\documentclass[10pt,a4paper,final]{iopart}
\expandafter\let\csname equation*\endcsname\relax
\expandafter\let\csname endequation*\endcsname\relax
\usepackage[margin=1 in]{geometry}
\usepackage{amsmath,color,iopams,graphicx}
\usepackage[breaklinks=true,colorlinks=true,linkcolor=blue,urlcolor=blue,citecolor=blue]{hyperref}

\begin{document}
\title{Non-chiral bosonization of strongly inhomogeneous Luttinger liquids}
\author{Joy Prakash Das, Chandramouli Chowdhury and Girish S. Setlur$^{*}$}
\address{Department of Physics \\ Indian Institute of Technology  Guwahati \\ Guwahati, Assam 781039, India}
\ead{$^{*}$gsetlur@iitg.ernet.in}

\begin{abstract}
Non-chiral bosonization (NCBT) is a non-trivial modification of the standard Fermi-Bose correspondence in one spatial dimensions made in order to facilitate the study of strongly inhomogeneous Luttinger liquids (LL) where the properties of free fermions plus the source of inhomogeneities are reproduced exactly. The formalism of NCBT is introduced and limiting case checks, fermion commutation rules, point splitting constraints, etc. are discussed. The Green functions obtained from NCBT are expanded in powers of the fermion-fermion interaction strength (forward scattering short-range only) and compared with the corresponding terms obtained using standard fermionic perturbation theory. Lastly, the Green functions obtained from NCBT are inserted into the Schwinger-Dyson equation which is the equation of motion of the Green functions and serves as a non-perturbative confirmation of the method. Some other analytical approaches like functional bosonization and numerical techniques like DMRG, which can be used to obtain the correlation functions in 1D,  are briefly discussed.
\end{abstract}


\vspace{2pc}
\section{ Introduction}

The study of strongly correlated electrons in one dimension has fascinated condensed matter physicists since the last several decades. The first soluble model of a strongly correlated Fermi system in one spatial dimension was given by Tomonaga \cite{tomonaga1950remarks} who was able to show that an assembly of Fermi particles in one dimension can be described by a quantized field of sound waves in the Fermi gas, where the sound waves obey Bose statistics. Later in 1963, essential achievements to this model was reflected in the work of Luttinger \cite{luttinger1963exactly} and a couple of years later, Mattis and Lieb provided an exact solution to this model \cite{mattis1965exact}. After that substantial progress has been made in understanding the properties of the 1D electron systems in the works of Dzyaloshinkii, Larkin, Efetov, etc. \cite{dzyaloshinskii1974correlation,efetov1976correlation} and later Haldane in his famous work \cite{haldane1981luttinger} developed the fundamentals of modern bosonization which essentially the Luther-Haldane operator approach. This method has been widely adopted to deal with correlations in a one dimensional fermionic system where the Fermi field operators are explicitly represented in terms of Bose operators. This method is well described in the book by Giamarchi \cite{giamarchi2004quantum}. This approach, also known as g-ology, is able to provide explicit expressions for the N-point functions  of a homogeneous or nearly homogeneous (where the inhomogeneities are over length scales large compared to the Fermi wavelength) Luttinger liquid \cite{haldane1981luttinger} in presence of forward scattering interactions. However this method fails when the translational symmetry is broken by  one or more delta function impurities or a step potential etc.\cite{setlur2013dynamics,das2018quantum}. On the other hand, the study of impurities in a quantum system constitutes a major theme in many body physics, and the 1D system is no exception \cite{kane1992transport} with a good number of practical realizations \cite{schwartz1998chain,bockrath1999luttinger,auslaender2000experimental}. To take into account strongly inhomogeneous  systems, a modification of the above method was suggested and a reconstruction of the field operator in terms of currents and densities is undertaken \cite{das2018quantum}. This approach, which goes by the name ``Non-chiral bosonization technique'', is based on a non-standard harmonic analysis of the rapidly varying parts of the density fields. It has been successfully   applied to yield the most singular part of the Green functions of different categories of systems like a cluster of impurities around an origin \cite{das2018quantum}, one step ladder \cite{das2017one}, one and two slowly moving impurities \cite{das2018ponderous}, etc.

Any novel technique needs validation. A usual way of doing this is to compare its results with those obtained using well established techniques. But the very fact that the same results can be obtained using other techniques would lessen the importance of a novel approach. In this regard, as already mentioned, the standard g-ology method is unable to produce the two-point functions of even free fermions with an impurity without resorting to renormalization methods. Another validation route for a new approach would be to adopt a perturbative approach and calculate the Green functions upto a certain order of the interaction term and compare with conventional fermionic perturbation theory. But the most convincing validation would be to insert the putative best Green functions obtained by any technique into the Schwinger-Dyson equation \cite{dyson1949s,schwinger1951green} which are the equations of motion of the Green functions and show that the result is an identity. This route is much more convincing than any other analytical or numerical comparison.

There is an alternative to the operator method of bosonization, which we also briefly discuss here is based on the idea that the Green functions of systems with mutual interaction between fermions may be expressed  as functional averages of the Green functions of free fermions interacting with arbitrary external potentials.
The averages are performed by treating the external potentials as Gaussian random variables with a mean determined by the actual external potentials present in the problem and the standard deviations related to the forward scattering strength of the short-range interaction between fermions. This approach, originally suggested by Fogedby \cite{fogedby1976correlation}, was elaborated by Lee and Chen \cite{lee1988functional}.  This approach was further expanded in \cite{fernandez2001friedel} and \cite{lerner2002strongly} where it was used to deal with impurities in a Luttinger liquid.
When it comes to numerical validation, the obvious tools that springs to mind are density-matrix renormalization group (DMRG) \cite{schollwock2005density}, finite-size scaling, etc. But performing numerical validation of an analytical result is somewhat like asking a high-school pupil to prove the analytical formulas for the solution of a quadratic equation by solving the latter numerically. The pupil would rightly argue that it is much more convincing and easier to simply insert the putative analytical solution back into the defining equation and show that the result is an identity. This is precisely what Schwinger Dyson validation does. Moreover for gapless systems such as the ones under consideration in the work, DMRG has its own shortcomings which is discussed in a later section.

In this work, it has been emphasized that inserting the Green functions into the Schwinger-Dyson equation and checking for an identity is superior to any other methods of validation. The next section describes the systems under study. Section 3 briefly describes the bosonization methods used to obtain the Green functions of the systems under study while section 4 illustrates the necessary validation checks the Green functions must obey. In section 5, the perturbative comparison is discussed whereas section 6 elaborates the Schwinger Dyson validation. The subsequent sections briefly describes the functional bosonization method and the DMRG method as to why they are not suitable in this regard.

\section{ System description}
The systems under study are Luttinger liquids with short ranged forward scattering mutual interactions between the fermions, the generic Hamiltonian being given by,
\begin{equation}
\begin{aligned}
H =& \int^{\infty}_{-\infty} dx \mbox{    } \psi^{\dagger}(x) \left( - \frac{1}{2m} \partial_x^2 + V(x) \right) \psi(x)
  + \frac{1}{2} \int^{ \infty}_{-\infty} dx \int^{\infty}_{-\infty} dx^{'} \mbox{  }v(x-x^{'}) \mbox{   }
 \rho(x) \rho(x^{'})
\label{Hamiltonian}
\end{aligned}
\end{equation}
The first term represents the kinetic energy while the second term represents the potential energy which is set to zero for homogeneous systems. For strongly inhomogeneous systems, it can be modeled as a finite sequence of barriers and wells clustered around a point (taken to be the origin, $x=0$). This potential cluster, which breaks the homogeneity of the system, can be as simple as a delta impurity $V_0\delta(x)$, two delta impurities placed close to each other $V_0( \delta(x+a)+\delta(x-a))$, finite barrier/well $\pm V \theta(x+a)\theta(a-x)$ and so on, where $\theta(x)$ is the Heaviside step function. The last term in equation (\ref{Hamiltonian}) represents the forward scattering mutual interaction and can be written as
\begin{equation}
\hspace{2 cm} v(x-x^{'}) = \frac{1}{L} \sum_{q}  v_q \mbox{ }e^{ -i q(x-x^{'}) }
\label{vq}
\end{equation}
where $ v_q = 0 $ if $ |q| > \Lambda $ for some fixed bandwidth $ \Lambda \ll k_F $ and $ v_q = v_0 $ is a constant, otherwise.
The systems are subjected to the RPA (Random Phase Approximation) to make analytical solutions feasible. In this limit, both the Fermi momentum and the mass of the fermion are allowed diverge keeping their ratio, viz., the Fermi velocity finite (i.e. $ k_F, m \rightarrow \infty $ but $ k_F/m = v_F < \infty  $), thereby linearizing the energy momentum dispersion near the Fermi surface ($E=E_F+p v_F$ instead of $E=p^2/(2m)$) \cite{stone1994bosonization}. Units are chosen such that $ \hbar = 1 $ and $ k_F $ is both the Fermi momentum as well as a wavenumber .

The central quantities that are used in the calculation of the Green functions are the transmission (T) and reflection (R) amplitudes of the non-interacting system plus the cluster of impurities which is easily calculated using elementary quantum mechanics. They are provided in a recent work \cite{das2018quantum}, which also contains the Green functions of such strongly inhomogeneous systems calculated using NCBT. On the other hand, the Green functions for the homogeneous case has been calculated much earlier by Mattis and Lieb \cite{mattis1965exact} and are explained for example, in the textbook by Giamarchi \cite{giamarchi2004quantum}. It is, of course, redundant to validate the well-established results of homogeneous LL. However this is done to demonstrate the validation method itself which is then used for the results obtained by the recently developed approach to bosonization - the NCBT \cite{das2018quantum}.

\section{ Bosonization and Green functions}
 In bosonization, a fermionic field is represented as an exponential of a bosonic field. This involves inverting the defining formulas for current and number densities viz. $ j(x) = Im[\psi^{\dagger}(x)\partial_x\psi(x)] $ and  $ \rho(x)  = \psi^{\dagger}(x)\psi(x) $ and rewriting $  \psi(x) $ in terms of $ j $ and $ \rho $. Then the continuity equation  $ \partial_t \rho + \partial_x j = 0 $ is invoked and $ \psi(x) $ is written purely as a (non-local) function of $ \rho $ and $ \partial_t \rho $. Thus  the N-point function is some combination of the correlations of the density field with itself.

The inversion of the defining relation between current and densities in the standard bosonization scheme that goes by the name g-ology  \cite{giamarchi2004quantum} yields the following relation between $ \psi_{\nu}(x,\sigma,t) $ (where $ \nu =$ R(+1) \mbox{  }or\mbox{   } L(-1)  for right and left movers respectively) and the slowly varying part of the density (this is a mnemonic for generating the N-point functions),
\begin{equation}
\begin{aligned}\hspace{-1.5in}
\psi_{\nu}(x,\sigma,t) \sim e^{ i \theta_{\nu}(x,\sigma,t) }
\label{PSINUg}
\end{aligned}
\end{equation}
with the local phase given by the formula,
\small
\begin{equation}
\begin{aligned}
 \theta_{\nu}(x,\sigma,t) = \pi \int^{x}_{sgn(x)\infty} dy \Big(& \nu  \mbox{  } \rho_s(y,\sigma,t) -  \int^{y}_{sgn(y)\infty} dy^{'} \mbox{ }\partial_{v_F t }  \mbox{ }\rho_s(y^{'},\sigma,t) \Big)
\label{localphase}
\end{aligned}
\end{equation}\normalsize
The above prescription in equation (\ref{PSINUg}) is valid only for homogeneous systems and for a half line (no tunneling across the barrier) and the Green functions in both cases are provided in \hyperref[AppendixA]{Appendix A}.

Analogous to conventional bosonization schemes \cite{giamarchi2004quantum}, the fermionic field operator in NCBT is also expressed in terms of currents and densities. But in NCBT, the field operator is modified to include the effect of back-scattering by impurities making it suitable to study translationally non-invariant systems such as the ones mentioned in the last section. The modified field operator of NCBT may be written as follows \cite{das2018quantum}.
\begin{equation}
\begin{aligned}
\psi_{\nu}(x,\sigma,t) \sim C_{\lambda  ,\nu,\gamma}\mbox{ }e^{ i \theta_{\nu}(x,\sigma,t) + 2 \pi i \lambda \nu  \int^{x}_{sgn(x)\infty}\mbox{ } \rho_s(-y,\sigma,t) dy}
\label{PSINU}
\end{aligned}
\end{equation}
Here $\theta_{\nu}$ is the familiar local phase given by equation (\ref{localphase}).
NCBT differs by the addition of the optional term $\rho_s(-y,\sigma,t)$ to this local phase that ensures the necessary trivial exponents for the single particle Green functions for a system of otherwise free fermions with impurities (which may also be obtained using standard Fermi algebra). The adjustable parameter $\lambda$ can take values either 0 or 1, which decides the presence or absence of the new term. In other words, setting $\lambda=0$ reduces the NCBT operator  to standard bosonization operator given in equation (\ref{PSINUg}). The factor $2 \pi i$ ensures that the field operator obeys the necessary fermion commutation rules since this term does not change the statistics of the field operator.
$C_{\lambda  ,\nu,\gamma}$ are pre-factors which are fixed by comparison using the non-interacting Green functions obtained from Fermi algebra.
 The field operator as given in equation (\ref{PSINU}) is to be treated as a mnemonic to obtain the Green functions rather than an operator identity, which avoids the necessity of the Klein factors that are conventionally used. The field operator (annihilation) is clubbed together with another such field operator (creation) and after fixing the C's and $\lambda$'s, one obtains the non-interacting two-point functions in terms of density-density correlation functions of the system. Lastly, these density correlation functions are replaced by their interacting versions to obtain the many-body Green functions for the strongly inhomogeneous LL under study are given in \hyperref[AppendixB]{Appendix B}. The details are described in an earlier work \cite{das2018quantum}.

\section{ Necessary validation checks}

\subsection{Commutation rules}
The Green functions under consideration in the present work have been studied using standard g-ology methods \cite{giamarchi2004quantum} and the recently developed NCBT \cite{das2018quantum} both of which are a field-theoretic approach to bosonization. In both these approaches,  the fermion field operator is expressed as a function of currents and densities. It is necessary that these operators obey the necessary commutation rules, which is the first mandatory step in validating the results.
\subsubsection{ Fermi Language }
\noindent {\bf{Fermi fields:}} Let there be $ N $ species of fermions $ \psi_j(x) $ where $ j = 1,2,.... , N $. By definition we have,
\begin{equation}
\{ \psi_j(x,t),  \psi_k(x^{'},t) \} = 0 \mbox{    } ; \mbox{  }
\{ \psi_j(x,t),  \psi^{\dagger}_k(x^{'},t) \} = \delta_{j,k} \mbox{   }\delta(x-x^{'})
\label{FERCOM}
\end{equation}
\\
\noindent {\bf{ Forward relation: }}  The currents and the densities are defined in terms of the fields as follows (no point splitting  etc. are needed in this general approach which makes no approximations at the outset of any sort - RPA or otherwise).
\begin{equation}
j_k(x,t) = Im[ \psi^{\dagger}_k(x,t)\partial_x\psi_k(x,t)] \mbox{   } ; \mbox{   } \rho_k(x,t) =  \psi^{\dagger}_k(x,t)\psi_k(x,t)
\label{DEF}
\end{equation}
\noindent {\bf{Current Algebra:}} The densities and the currents obey current algebra.
\begin{equation}
\begin{aligned}
&[\rho_k(x,t), \rho_l(x^{'},t)] = 0 \\
&[\rho_k(x,t), j_l(x^{'},t)] = i\mbox{     }  \delta_{l,k}\mbox{  }\rho_l(x^{'},t)\partial_{x^{'} } \delta(x-x^{'})\\
&[j_k(x,t),j_l(x^{'},t)] = -i\mbox{  } \delta_{k,l} \mbox{  }j_l(x^{'},t) \partial_x \delta(x-x^{'}) + i\mbox{  } \delta_{k,l} \mbox{  }  j_k(x,t) \partial_{x^{'}} \delta(x-x^{'})
\label{CURRALG}
\end{aligned}
\end{equation}
\noindent{\bf{Field-Current/Density commutators:}} The fields and the currents/densities   obey the following equal-time commutation rules.
\begin{equation}
\begin{aligned}
& [\psi_k(x,t), \rho_l(x^{'},t)] = \delta_{k,l} \mbox{   }\delta(x-x^{'}) \mbox{  }\psi_k(x,t)  ;\\
&[\psi_k(x,t), j_l(x^{'},t)] = \frac{1}{2i} (\delta_{k,l} \delta(x-x^{'})\mbox{  } (\partial_{x^{'}} \psi_l(x^{'},t)) - \delta_{k,l}
 (\partial_{x^{'}} \delta(x-x^{'})) \psi_l(x^{'},t) )
 \label{FIELDCURR}
 \end{aligned}
\end{equation}\normalsize

\subsubsection{ Bose Language }

\mbox{   }

\noindent {\bf{Boson Fields:}} Define self adjoint $ \pi_j(x,t) $ and $ \rho_j(x,t) $, $ j = 1,2,3, ... , N $ obeying canonical commutation rules.
\begin{equation}
\begin{aligned}
.[ \rho_j(x,t),  \rho_k(x^{'},t) ] =  0 \mbox{    } \mbox{    } ; \mbox{  }[ \pi_j(x,t),  \pi_k(x^{'},t) ] = 0\mbox{ } ;\mbox{ }
 [ \pi_j(x,t),  \rho_k(x^{'},t) ] = i\delta_{j,k} \mbox{   }\delta(x-x^{'})
\label{BOSECOM}
\end{aligned}
\end{equation}
\noindent {\bf{Forward relation: }}
\begin{equation}
j_k(x,t) = - \rho_k(x,t) \partial_x \pi_k(x,t)
\label{CURR}
\end{equation}
Equation (\ref{CURR}) together with equation (\ref{CURRALG}) implies  equation (\ref{BOSECOM})  provided $ \rho_k(x,t) $  does not vanish anywhere since division by this quantity is needed.
\\

\noindent {\bf{Conjecture:}} Fermi-Bose Correspondence:
\begin{equation}
\begin{aligned}
\psi_k(x,t) =& e^{ i \pi \sum_{ l < k } \int^{\infty}_{ -\infty } dy \mbox{  } \rho_l(y,t) }  \frac{1}{ \sqrt{ N^{0} } }\sum_{p} n_F(p) \mbox{  }
e^{ i \xi(p) }\mbox{   }
e^{ i \pi sgn(p)\int^{x}_{ sgn(x)\infty } dy \mbox{  } \rho_k(y,t) } e^{ -i \pi_k(x,t) } \mbox{   }\sqrt{ \rho_k(x,t) }
\label{FERBOSE}
\end{aligned}
\end{equation}
where $ n_F(p) = \theta(k_F-|p|) $ and $ N^{0 } = \sum_{p} n_F(p) $. The equation (\ref{FERBOSE}) inserted into equation (\ref{DEF}) leads to an identity together with equation (\ref{CURR}) provided the following identification is made,
$  e^{ i \xi(p) }\mbox{   } e^{ -i \xi(p^{'}) } = \delta_{p,p^{'}} $ (imagine e.g. , $ \xi(p) \neq \xi(p') $  when $ p \neq p' $ to be a real quantity that tends to infinity for all $ p $). \\ \mbox{  }  \\

\noindent {\bf{Theorem:}} The conjecture in equation (\ref{FERBOSE}) obeys fermion commutation rules in equation (\ref{FERCOM}) in conjunction with equation (\ref{BOSECOM}) and $ e^{ i \xi(p) }\mbox{   } e^{ -i \xi(p^{'}) } = \delta_{p,p^{'}} $. \\

\noindent The proof of the above theorem is given in \hyperref[AppendixC]{Appendix C}. The central NCBT relation  between  the slow part of the Fermi field and  current/densities viz. equation (\ref{PSINU}) may be obtained from the conjecture in equation (\ref{FERBOSE}) using harmonic analysis of the density operator, which goes as follows.
\begin{equation}
\rho(x,\sigma,t) = \rho_0+\rho_s(x,\sigma,t) + e^{ 2 i k_F x } \mbox{   }\rho_f(x,\sigma,t) +  e^{ - 2 i k_F x } \mbox{   }\rho^{*}_f(x,\sigma,t)
\label{Harmonic}
\end{equation}
Here $\rho_s$ is the slow part and $\rho_f$ is the oscillating part of the density. According to Haldane's harmonic analysis \cite{haldane1981luttinger}, the fast part can be expressed in terms of the slow part as follows.
\begin{equation}
\hspace{1.5 cm}\rho_f(x,\sigma,t)  \sim e^{ 2 i \pi \int^{x}_{-\infty} \rho_s(y,\sigma,t) dy }
\label{Haldane}
\end{equation}
Using Haldane's harmonic analysis in equation (\ref{Harmonic}) and inserting to the conjecture in equation (\ref{FERBOSE}) and extracting the slow part, one can obtain the expression of the field operator used in standard bosonization, given by equation (\ref{PSINUg}). On the other hand, NCBT uses a non-standard harmonic analysis which is ideally suited to study systems with a cluster of impurities \cite{das2018quantum}. Non-standard harmonic analysis means the replacement,
\begin{equation}
\hspace{1.5 cm}\rho_f(x,\sigma,t)  \sim e^{ 2 i \pi \int^{x}_{-\infty} (\rho_s(y,\sigma,t) + \lambda \rho_s(-y,\sigma,t)) dy }
\label{NCBTHH}
\end{equation}
Using this non-standard harmonic analysis in equation (\ref{Harmonic}) and inserting it to the conjecture in equation (\ref{FERBOSE}) and extracting the slow part, one can obtain the expression of the field operator used in non chiral bosonization, given by equation (\ref{PSINU}). Note that the variable $\lambda$ in equation (\ref{NCBTHH}) can take values only 0 or 1, as also discussed in the last section and the the non-standard harmonic analysis reduces to the standard one when $\lambda=0$.\\\\

\noindent{\bf Field-Current/Density commutators:} Lastly, the identities in equation (\ref{FIELDCURR2}) below are obeyed regardless of whether these commutators are evaluated in the usual Fermi language or using the conjecture  equation (\ref{FERBOSE})  and  the canonical commutators  equation (\ref{BOSECOM}).  However division by $ \rho_k(x) $ should be permissible to prove the same.
\begin{equation}
\begin{aligned}
 &[\psi_k(x,t), \rho_l(x^{'},t)] = \delta_{k,l} \mbox{   }\delta(x-x^{'}) \mbox{  }\psi_k(x,t)  ;
 \\&[\psi_k(x,t), j_l(x^{'},t)] =
 \frac{1}{2i} (\delta_{k,l} \delta(x-x^{'})\mbox{  } (\partial_{x^{'}} \psi_l(x^{'},t)) - \delta_{k,l}
 (\partial_{x^{'}} \delta(x-x^{'})) \psi_l(x^{'},t) )
 \label{FIELDCURR2}
 \end{aligned}
\end{equation}
It must be stressed that the addition of the extra term $\rho_s(-y,\sigma,t)$ in equation (\ref{PSINU}) of the NCBT scheme does not violate any of the commutation rules above because of the constant $2\pi i$ in it and the integration of the density is just a natural number.  It is also to be noted that there are some additional global quantities that are needed to be incorporated into the C-numbers in equation (\ref{PSINU}) to make sure the up-spin field anti-commutes with the down spin field, the left leg of the ladder anti-commutes with the right leg (in case of a one step fermionic ladder) and so on. They are described in an earlier work \cite{das2017one}.

\subsection{Limiting case checks}
Another necessary criterion to be obeyed by the correlation functions are the limiting case checks. For the homogeneous case, there is just one limiting case, viz., switching off mutual interactions between the particles. Under this condition the LL Green functions must reduce to the free particle Green functions. In absence of interactions, the holon velocity becomes equal to the Fermi velocity ($v_h \to v_F$) and thus the Green functions in equation (\ref{HSS}) takes the form of that of a free particle.
\begin{equation*}\footnotesize
\begin{aligned}
\Big\langle T\mbox{  }\psi_{R}(x_1,\sigma,t_1 )\psi_{R}^{\dagger}(x_2,\sigma,t_2)\Big\rangle \sim
\frac{1}{ (x_1-x_2 -v_F (t_1-t_2))} \mbox{ };\mbox{ }
\Big\langle T\mbox{  }\psi_{L}(x_1,\sigma,t_1)\psi_{L}^{\dagger}(x_2,\sigma,t_2)\Big\rangle \sim
\frac{1}{(-x_1+x_2 -v_F (t_1-t_2))} \\
\end{aligned}
\end{equation*}\normalsize
On the other hand, for the strongly inhomogeneous systems, the Green functions obtained using NCBT can be subjected to various limiting cases as follows. \\

\noindent{\bf No interaction. }
By switching off the inter-particle interactions between particles ( $v_0 = 0$ ), one obtains the Green functions for free fermions plus impurities which can also be obtained using Fermi algebra. In such a case, the holon velocity is equal to the Fermi velocity ($v_h \to v_F$)  and equation (\ref{SS}) and equation (\ref{OS}) will take the form of single particle Green functions of such inhomogeneous systems.\\

\noindent{\bf No impurity. }
In absence of any impurity, there is no reflection ($|R|=0$) and there is no concept of opposite sides of the origin as its a homogeneous case. There will be no reflection  terms such as $\langle\psi_{R}\psi_{L}^{\dagger}\rangle$ and $\langle\psi_{L}\psi_{R}^{\dagger}\rangle$ in this case. The only non-zero terms are the transmission propagators $\langle\psi_{R}\psi_{R}^{\dagger}\rangle$ and $\langle\psi_{L}\psi_{L}^{\dagger}\rangle$ whose exponents takes the following form:\small
\begin{equation*}
P=\frac{(v_h+v_F)^2}{8v_hv_F};\mbox{ }Q=\frac{(v_h-v_F)^2}{8v_hv_F};\mbox{ }X=\gamma_1=0;
\end{equation*}
\normalsize
Using the above, one obtains the precise Green functions of the standard homogeneous Luttinger liquid as given in equation (\ref{HSS}).\\

\noindent{\bf No tunneling. }
For an infinite barrier ($|R|=1$), there is no need to consider the two points to be on the opposite sides of the impurity. While the Green functions for points on the same side of the origin as given in equation (\ref{SS}) takes the form that of a half line as given in equation (\ref{HL}). Also in this case, the full Green function vanishes when one of the points is at the location of the infinite barrier.\\

\noindent{\bf Far from impurity. }
Lastly it can be observed that when both the points are situated far away from the impurity and on the same side of it, then the transmission propagators $\langle\psi_{R}\psi_{R}^{\dagger}\rangle$ and $\langle\psi_{L}\psi_{L}^{\dagger}\rangle$ become  immune to the presence of impurities and takes the form of the homogeneous case (equation (\ref{HSS})). But the reflection  terms viz. $\langle\psi_{R}\psi_{L}^{\dagger}\rangle$ and $\langle\psi_{L}\psi_{R}^{\dagger}\rangle$ continue to be affected by the presence of the impurity since in these cases the region where the impurity is present needs to be traversed (in order for reflection to take place).\\

\subsection{Point splitting constraint}
It is mandatory that the field operators as in equation (\ref{PSINUg}) and equation (\ref{PSINU}) do not violate the point splitting constraints, which  is a crucial self-consistency check. Point splitting constraint is the assertion that the 
NCBT Green functions are consistent with current algebra. The use of the non-local expression for the field gives back the currents and densities which were used to exponentiate the commutation rules and write down the non-local expression in the first place.
\small
\begin{equation}
\begin{aligned}
\lim_{a \rightarrow 0 }\mbox{   }
(\psi^{\dagger}_{\nu}(x,\sigma,t)\psi_{\nu}(x+a,\sigma,t) - < \psi^{\dagger}_{\nu}(x,\sigma,t)\psi_{\nu}(x+a,\sigma,t)  >) =
\frac{1}{   2\nu  }  ( \nu  \mbox{  } \rho_s(x,\sigma,t) - \int^{x}_{sgn(x)\infty} dy^{'} \partial_{v_F t }  \rho_s(y^{'},\sigma,t) )
\label{PSC}
\end{aligned}
\end{equation}
\normalsize
 It may be seen below that it leads to constraints on the form for the product of the coefficient $ C_{\lambda  ,\nu,\gamma} $ in equation (\ref{PSINU}) and its complex conjugate which is to be regarded as one single object rather than a product of two complex numbers as the non-local expression for the field is merely a mnemonic - not be be taken literally. After some straightforward algebra which is given briefly below, the following constraints emerge from equation (\ref{PSC}) (here $ C_{\lambda  ,\nu}(x,\sigma) \equiv \sum_{ \gamma = \pm 1 } \theta(\gamma x) C_{\lambda  ,\nu,\gamma}(\sigma) $),\small
\begin{equation*}
\begin{aligned}
\lim_{a \rightarrow 0 }
\sum_{ \lambda,\lambda^{'} \in \{0,1\} }& <C^{\dagger}_{\lambda^{'} ,\nu}(x,\sigma) C_{\lambda  ,\nu}(x+a,\sigma)> e^{ \frac{1}{2} [Q_{\nu }(x,\sigma;\lambda^{'}), -Q_{\nu }(x+a,\sigma;\lambda) ]  }\\
\times&\left(e^{Q_{\nu }(x,\sigma;\lambda^{'}) -Q_{\nu }(x+a,\sigma;\lambda) } - <e^{Q_{\nu }(x,\sigma;\lambda^{'}) -Q_{\nu }(x+a,\sigma;\lambda) } > \right )
 =\frac{1}{   2\nu  }  ( \nu  \mbox{  } \rho_s(x,\sigma,t) - \int^{x}_{sgn(x)\infty} dy^{'} \partial_{v_F t }  \rho_s(y^{'},\sigma,t) )
\end{aligned}
\end{equation*}
\normalsize
where $Q_{\nu }(x,\sigma;\lambda^{'}) =  -i \theta_{\nu }(x,\sigma,t) - 2 \pi i \nu \lambda^{'} \int^{x}_{sgn(x)\infty} \rho_s(-y,\sigma,t) \mbox{  }dy
 $.\\\\When $ \lambda $ and $ \lambda^{'} $ are unequal,
\begin{equation*}\footnotesize
\begin{aligned}
<C^{\dagger}_{1,\nu}(x,\sigma) C_{0,\nu}(x,\sigma)>
\lim\limits_{a \rightarrow 0 }&\mbox{   }e^{ \frac{1}{2} [Q_{\nu }(x,\sigma;1), -Q_{\nu }(x+a,\sigma;0) ]  }
 =<C^{\dagger}_{0,\nu}(x,\sigma) C_{1 ,\nu}(x,\sigma)>\lim\limits_{a \rightarrow 0 }\mbox{   }e^{ \frac{1}{2} [Q_{\nu }(x,\sigma;0), -Q_{\nu }(x+a,\sigma;1) ]  }
\end{aligned}
\end{equation*}\normalsize
where $
[Q_{\nu }(x,\sigma;\lambda^{'}), -Q_{\nu }(x+a,\sigma;\lambda) ]
 = - \pi \mbox{  } \nu  i  - 2 \pi  \nu \lambda  i \theta(2x+a)
 + 2 \pi  \nu \lambda^{'}
i\theta(2x+a) $.\\\\
When they are equal however, the following results emerge,\footnotesize
\[
\lim\limits_{a \rightarrow 0 }\mbox{   }
\sum_{ \lambda \in \{0,1\} }\mbox{  } <C^{\dagger}_{\lambda,\nu}(x,\sigma) C_{\lambda  ,\nu}(x+a,\sigma)>
e^{ \frac{1}{2} ( - \pi \mbox{  } \nu  i ) }
\mbox{  }
(- a) \mbox{  } \partial_x Q_{\nu }(x,\sigma;\lambda)
 =
\frac{1}{   2\nu  }  ( \nu  \mbox{  } \rho_s(x,\sigma,t) - \int^{x}_{sgn(x)\infty} dy^{'} \partial_{v_F t }  \rho_s(y^{'},\sigma,t) )
\]\normalsize
Note that $
\partial_x Q_{\nu }(x,\sigma;\lambda) =  -i \partial_x\theta_{\nu }(x,\sigma,t) - 2 \pi i \nu \lambda \rho_s(-x,\sigma,t) $ hence,
\[
\lim\limits_{a \rightarrow 0 }\mbox{   }
\sum_{ \lambda \in \{0,1\} }\mbox{  } <C^{\dagger}_{\lambda,\nu}(x,\sigma) C_{\lambda  ,\nu}(x+a,\sigma)>
e^{ \frac{1}{2} ( - \pi \mbox{  } \nu  i ) }
\mbox{  }
(- a) \mbox{  } ( - 2 \pi i \nu \lambda \rho_s(-x,\sigma,t))
 =
0
\]
This means,
\[
 <C^{\dagger}_{0,\nu}(x,\sigma) C_{0,\nu}(x+a,\sigma)>
 =
\frac{e^{ \frac{i}{2}  \pi \mbox{  } \nu   }
}{   2\pi i \nu  a  }  = \frac{1}{2\pi a}
\]
and,
\[
<C^{\dagger}_{1,\nu}(x,\sigma) C_{1,\nu}(x+a,\sigma)> = 0
\]
These may be compactly written as follows. Define
\[
<C^{\dagger}_{\lambda^{'} ,\nu^{'}}(x^{'},\sigma^{'}) C_{\lambda  ,\nu}(x,\sigma)>
\rightarrow C_2(\lambda^{'} ,\nu^{'},x^{'},\sigma^{'}; \lambda  ,\nu,x,\sigma)
\]
so the final point splitting constraints take the following form,
\begin{equation}
\begin{aligned}
&C_2(0,\nu,x,\sigma; 0,\nu,x+a,\sigma)  = \frac{1}{2\pi a}\\
&C_2(1,\nu,x,\sigma;0,\nu,x+a,\sigma) =  C_2(0,\nu,x,\sigma; 1,\nu,x+a,\sigma)\\
&C_2(1,\nu,x,\sigma; 1,\nu,x+a,\sigma) = 0\\
\end{aligned}
\end{equation}\normalsize
 This is the reason why, in the evaluation of the two-point function, the possibility of both $ \lambda = \lambda^{'} = 1 $  was never considered (the corresponding C's are zero). Thus the NCBT formulas for the Green functions obey the point splitting constraints (so does   standard bosonization).
\section{ Perturbative comparison}

The Green functions of a homogeneous Luttinger liquid obtained using g-ology method (given in \hyperref[AppendixA]{Appendix A}) can be verified by a comparison with those obtained using standard fermionic perturbation. For this, the Green functions are expanded in powers of the interaction parameter $v_0$ (see equation (\ref{vq})). Note that the holon velocity $v_h$ is related to the Fermi velocity and the interaction parameter $v_0$ by the relation $v_h=v_F\sqrt{1+2v_0/(\pi v_F)}$. On the other hand, the zeroth and the first order terms of the perturbation series can be calculated as follows ({\bf Notation:} $X_i \equiv (x_i,\sigma_i,t_i)$).\\
\begin{equation}
\begin{aligned}
&\delta \langle \psi_{\nu_1}(X_1) \psi^{\dagger}_{\nu_2}(X_2) \rangle^0 = \langle T \psi_{\nu_1}(X_1) \psi^{\dagger}_{\nu_2}(X_2) \rangle_0\\
&\delta \langle \psi_{\nu_1}(X_1) \psi^{\dagger}_{\nu_2}(X_2) \rangle^1 = - \frac{i}{2}
 \int d\tau \mbox{   } \int dy \int dy^{'} \mbox{   }v(y-y^{'})  \langle T \mbox{  } \rho_s(y,\tau_{+})  \rho_s(y^{'},\tau) \psi_{\nu_1}(X_1) \psi^{\dagger}_{\nu_2}(X_2) \rangle_0
\label{pert}
\end{aligned}
\end{equation}
Here $v(y-y') = v_0\delta(y-y')$ is the short ranged mutual interaction term, the $\nu$'s represent either a right mover ($\nu_i = 1$) or a left mover ($\nu_i=-1$) and $\rho_s = \psi^{\dagger}_R\psi_R + \psi^{\dagger}_L\psi_L$ is the slow part of the density function. The symbol $\langle..\rangle_0$ on the RHS indicates single particle functions.
Using equation (\ref{pert}), the Green functions of a homogeneous LL up to the first order in interaction parameter can be obtained as follows:
\begin{equation*}
\begin{aligned}
G_{RR}&(x_1,x_2,t_1-t_2)
=& \frac{i}{2\pi}\frac{1}{(x_1-x_2)-v_F(t_1-t_2)}+ \frac{i (t_1-t_2)}{4 \pi^2 ((x_1-x_2)-v_F(t_1-t_2))^2}v_0
\end{aligned}
\end{equation*}
This is precisely the same as that obtained by expanding equation (\ref{HSS}) obtained using standard bosonization in powers of $v_0$ and keeping up to the first order.

The Green functions obtained from NCBT technique (given in \hyperref[AppendixB]{Appendix B}) are claimed to be the most singular part of the full Green function \cite{das2018quantum}. Similar to the case above, the Green functions of  strongly inhomogeneous systems (with reflection amplitude R and transmission amplitude T) are calculated perturbatively using equation (\ref{pert}), while treating the source  of inhomogeneities exactly. After retaining the most singular terms in the first order (the zeroth order is a single term), the following results are obtained (the {\bf R} and {\bf T} on the RHS are reflection and transmission amplitudes respectively).\\

\noindent{\bf{ $ G_{RR} $ : $ x_1,x_2 > 0 $ }}
\begin{equation*}
\begin{aligned}
G_{RR}&(x_1,x_2,t_1-t_2)
=& \frac{i}{2\pi}\frac{1}{(x_1-x_2)-v_F(t_1-t_2)}+ \frac{i (t_1-t_2)}{4 \pi^2 ((x_1-x_2)-v_F(t_1-t_2))^2}v_0
\end{aligned}
\end{equation*}
{\bf{ $ G_{RL} $ : $ x_1,x_2 > 0 $ }}
\begin{equation*}
\begin{aligned}
G_{RL}&(x_1,x_2,t_1-t_2)
 =& \frac{i{\bf R}}{2\pi}\frac{1}{(x_1+x_2)-v_F(t_1-t_2)}+  \frac{ i {\bf R} (t_1-t_2)}{4\pi^2((x_1+x_2)-v_F(t_1-t_2))^2}v_0
\end{aligned}
\end{equation*}
{\bf{ $ G_{RR} $ : $ x_1>0,x_2 < 0 $ }}
\begin{equation*}
\begin{aligned}
G_{RR}&(x_1,x_2,t_1-t_2)
=&\frac{i{\bf T}}{2\pi}\frac{1}{(x_1-x_2)-v_F(t_1-t_2)}+ \frac{i{\bf T} (t_1-t_2)}{4 \pi^2 ((x_1-x_2)-v_F(t_1-t_2))^2}v_0
\end{aligned}
\end{equation*}
These are precisely the same as those obtained by expanding equation (\ref{SS}) and equation (\ref{OS}) obtained using NCBT in powers of $v_0$ and retaining up to the first order and also discarding the less singular terms at this order.
The notion of ``the most singular part" of an expression may be made sense of in the following manner. Think of these as function of the time difference $ \tau = t-t' $ (which they are). In the formulas that are encountered while expanding in powers of the coupling, there are going to be terms of the form (e.g.)
\[
\frac{ A \mbox{  }\tau }{ (\tau - a)^2 } + \frac{B}{(\tau-a_1)(\tau-a_2)}
\]
The first term is regarded as more singular than the second (if $ a_1 \neq a_2 $) since the former is a second order pole whereas the latter when partial fraction expanded are a sum of two first order poles. In the perturbative expansion of the single-particle Green function, pretending that Wick's theorem applies at the level of the density fluctuations is tantamount to retaining second order poles and discarding poles of a lower order.

\section{ Schwinger-Dyson equation}
The Schwinger-Dyson equation relates the two-point Green functions to certain four-point functions as follows:
\begin{equation}
\begin{aligned}
 G^{full}&_{\nu,\nu'}(x,x';t-t')=G^{0}_{\nu,\nu'}(x,x';t-t')\\
 -&i\sum_{\nu_1} \int dx_1\int dt_1 \mbox{ }G^{0}_{\nu,\nu_1}(x,x_1;t-t_1)   \int dy\mbox{ } v(x_1-y)
<T\mbox{  } \rho(y,t_{1} ) \psi_{\nu_1 }(x_1,\sigma ,t_1)\psi^{\dagger}_{\nu'}(x',\sigma,t') >_{full}\\
\end{aligned}
\end{equation}
where $\rho(y,t_{1} )= \rho(y,\uparrow,t_{1} ) +  \rho(y,\downarrow,t_{1} ) $ is the total density (sum of up spin and down spin density) and $v(y-y')$ is the mutual interaction. After operating the equation by $(i\partial_t + i \nu  v_F   \partial_x) $ and $(-i\partial_{t'} - i\nu'  v_F   \partial_{x'})$ we get  (only for $  x,x' \neq 0 $ and $ x \neq \pm x' $ and $ t \neq t' $)
\begin{equation}
\begin{aligned}
&(i\partial_t + i \nu v_F   \partial_x) G^{full}_{\nu,\nu'}(x,x';t-t')
=   \int dy \mbox{  }v(x-y)\mbox{  }   <T\mbox{  } \rho(y,t ) \psi_{\nu}(x,\sigma ,t)\psi^{\dagger}_{\nu'}(x',\sigma,t') >_{full}
\\
&(-i\partial_{t'} -i \nu'   v_F   \partial_{x'}) G^{full}_{\nu,\nu'}(x,x';t-t')
=   \int dy \mbox{  }v(x'-y) \mbox{  }  <T\rho(y,t' ) \psi_{\nu}(x,\sigma ,t)\psi^{\dagger}_{\nu'}(x',\sigma,t') >_{full}
 \label{dyson2}
\end{aligned}
\end{equation}
\normalsize
The next task is to examine whether the Green functions obtained using g-ology and those obtained using NCBT obey the above set of equations.
\subsection{Homogeneous case}
The Green functions for homogeneous LL can be calculated using the standard bosonization technique (equation (\ref{PSINUg})) where the local phase is given by equation (\ref{localphase}) as follows.

\begin{equation}
\begin{aligned}
\theta_{\nu}&(x,\sigma,t)=\pi \int^{x}_{sgn(x)\infty} dy\mbox{  } \Big(\nu \mbox{  }\rho_s(y,\sigma,t)
 - \int^{y}_{sgn(y)\infty} dy' \mbox{  }\partial_{ v_F t}\mbox{ } \rho_s(y,\sigma,t)  \Big)  \\
 \label{smalls}
\end{aligned}
\end{equation}
Using standard bosonization one can write
\begin{equation}
\begin{aligned}
&\psi_{\nu}(x,\sigma ,t) \rightarrow e^{ i \theta_{\nu}(x,\sigma ,t)   } \mbox{ }\mbox{ };\mbox{ }\mbox{ }
\psi_{\nu' }(x',\sigma ,t') \rightarrow e^{ i \theta_{\nu'}(x',\sigma ,t')   }\\
\end{aligned}
\end{equation}
Now,
\[
\lim\limits_{\epsilon \to 0} \frac{e^{\epsilon \rho} - 1}{\epsilon}=\rho.
\]
Hence the four point function in equation (\ref{dyson2}) can be written as
\begin{equation*}
\begin{aligned}
<\rho (y,t) \mbox{ }\psi_{\nu} (x,\sigma,t) \mbox{ } \psi_{\nu'}^{\dagger}(x',\sigma,t')>
=&\lim\limits_{\epsilon \to 0}<\frac{e^{\epsilon \rho(y,t)} - 1}{\epsilon}\mbox{ }e^{i \theta_{\nu} (x,\sigma,t)} \mbox{ } e^{-i \theta_{\nu'} (x',\sigma,t')} >\\
=&\lim\limits_{\epsilon \to 0}\frac{1}{\epsilon}< e^{\epsilon \rho(y,t) }\mbox{ }e^{i \theta_{\nu} (x,\sigma,t)} \mbox{ } e^{-i \theta_{\nu'} (x',\sigma,t')}
-e^{i \theta_{\nu} (x,\sigma,t)} \mbox{ } e^{-i \theta_{\nu'} (x',\sigma,t')}>\\
\end{aligned}
\end{equation*}
Choose,
\begin{equation*}
\begin{aligned}
\mathcal{A}=\mbox{ }\epsilon \rho(y,t)\mbox{ };\mbox{ }
\mathcal{B}=i \theta_{\nu} (x,\sigma,t)\mbox{ };\mbox{ }
\mathcal{C}=-i\theta_{\nu'} (x',\sigma,t')
\end{aligned}
\end{equation*}
Using Baker-Campbell-Hausdorff formula,
\[
<e^\mathcal{A} e^\mathcal{B} e^\mathcal{C}>= e^{\frac{1}{2} \mathcal{A}^2}e^{\frac{1}{2} \mathcal{B}^2}e^{\frac{1}{2} \mathcal{C}^2}e^{<\mathcal{A}\mathcal{B}>}e^{<\mathcal{B}\mathcal{C}>}e^{<\mathcal{A}\mathcal{C}>}
\]
Now $e^{\frac{1}{2} \mathcal{B}^2}e^{\frac{1}{2} \mathcal{C}^2}e^{<\mathcal{B}\mathcal{C}>}$ is the Green functions $<\psi_{\nu} (x,\sigma,t)\mbox{ } \psi_{\nu'}^{\dagger}(x',\sigma,t')>$ and $e^{\frac{1}{2} \mathcal{A}^2}$ can be ignored as $\epsilon$ is tending to zero. Hence we can write,
\begin{equation*}
\begin{aligned}
<\rho (y,t) \mbox{ }\psi_{\nu}& (x,\sigma,t) \mbox{ } \psi_{\nu'}^{\dagger}(x',\sigma,t')>\\
=&\lim\limits_{\epsilon \to 0}\bigg[
e^{< i \epsilon \rho(y,t) \mbox{ }\theta_{\nu}(x,\sigma,t)>} \mbox{ }
e^{<-i \epsilon \rho(y,t) \mbox{ }\theta_{\nu'}(x',\sigma,t')>}
-1\bigg]\frac{<\psi_{\nu} (x,\sigma,t)\mbox{ } \psi_{\nu'}^{\dagger}(x',\sigma,t')>}{\epsilon}\\
=&\lim\limits_{\epsilon \to 0}\bigg[
1+< i \mbox{ } \epsilon \mbox{ } \rho(y,t) \theta_{\nu}(x,\sigma,t)>
-< i \mbox{ } \epsilon  \mbox{ }\rho(y,t) \theta_{\nu'}(x',\sigma,t')>-1\bigg]
\frac{<\psi_{\nu} (x,\sigma,t)\mbox{ } \psi_{\nu'}^{\dagger}(x',\sigma,t')>}{\epsilon}\\
=&\mbox{ }\mbox{ }\bigg[< i \mbox{ }\rho(y,t) \theta_{\nu}(x,\sigma,t)>
-< i \mbox{ } \rho(y,t) \theta_{\nu'}(x',\sigma,t')>\bigg]
<\psi_{\nu}(x,\sigma,t)\mbox{ } \psi_{\nu'}^{\dagger}(x',\sigma,t')>
\end{aligned}
\end{equation*}
Using equation (\ref{smalls}),
\small
\begin{equation*}
\begin{aligned}
<\rho (y,t) \mbox{ }\psi_{\nu} (x,\sigma,t)& \mbox{ } \psi_{\nu'}^{\dagger}(x',\sigma,t')>\\
=&\bigg(< i \pi \nu\mbox{ } \int_{sgn(x)\infty}^{x} dz \mbox{ } \rho_h(y,t)\rho_h(z,t)
-i\pi\mbox{ }\int_{sgn(x)\infty}^{x} dz \mbox{ }\partial_{v_Ft} \int_{sgn(y)\infty}^z  \rho_h(y,t)\rho_h(z',t)dz' \mbox{ }>\\
-&< i \pi \nu'\mbox{ } \int_{sgn(x')\infty}^{x'} dz \mbox{ } \rho_h(y,t)\rho_h(z,t')
-i\pi\mbox{ }\int_{sgn(x')\infty}^{x'} dz \mbox{ }\partial_{v_Ft'} \int_{sgn(y)\infty}^z  \rho_h(y,t)\rho_h(z',t')dz' \mbox{ }>\Bigg)\\
& <\psi_{\nu}(x,\sigma,t)\mbox{ } \psi_{\nu'}^{\dagger}(x',\sigma,t')>
\end{aligned}
\end{equation*}\normalsize
Using the density density correlation functions from equation (\ref{RHOSRHOSg}), the following form of the Schwinger Dyson equation is obtained.
\begin{equation}
\begin{aligned}
(i\partial_t + i  v_F   \partial_x)  <T\mbox{  } \psi_{R}(x,\sigma ,t)\psi^{\dagger}_{R}(x',\sigma,t') >
=&v_0 \frac{i}{4 \pi   v_h^2}  \mbox{   }\left( \frac{v_h (v_h-v_F)}{-v_h(t-t')-x+x'}-\frac{v_h (v_F+v_h)}{v_h(t-t')-x+x'} \right)\\
&\mbox{   }\times <T\mbox{  } \psi_{R}(x,\sigma ,t)\psi^{\dagger}_{R}(x',\sigma,t') >
\label{RR}
\end{aligned}
\end{equation}\normalsize
Substituting the two point function from equation (\ref{HSS}),
\begin{equation*}
\begin{aligned}
-i &\bigg(\frac{P (v_h-v_F)}{v_h(t-t')-x+x'}+\frac{Q (v_F+v_h)}{v_h(t-t')+x-x'}\bigg)
=v_0 \frac{i}{4 \pi   v_h^2}  \mbox{   }(\frac{v_h (v_h-v_F)}{-v_h(t-t')-x+x'}-\frac{v_h (v_F+v_h)}{v_h(t-t')-x+x'} )
\end{aligned}
\end{equation*}
\normalsize
Upon inserting the explicit expressions of the anomalous exponents from  Eq (\ref{expH}) into the above equation, one obtains an identity. Thus the correlation functions obtained by g-ology methods satisfy the exact Schwinger Dyson equation, which is a non-perturbative validation of the same.
\subsection{Inhomogeneous case}

The Green functions of a Luttinger liquid with a cluster of impurities around a point can be calculated using the powerful  non-chiral bosonization technique (equation (\ref{PSINU})) where the familiar local phase undergoes modification and is given by the following ($\lambda=0,1$).
\begin{equation}
\begin{aligned}
\Theta_{\nu}&(x,\sigma,t,\lambda)=\pi \int^{x}_{sgn(x)\infty} dy\mbox{  } \Big(\nu \mbox{  }\rho_s(y,\sigma,t)
 - \int^{y}_{sgn(y)\infty} dy' \mbox{  }\partial_{ v_F t}\mbox{ } \rho_s(y,\sigma,t)  \Big) + 2 \pi \nu \lambda \int^{x}_{sgn(x)\infty}dy\mbox{   }\rho(-y,\sigma,t) \\
 \label{caps}
\end{aligned}
\end{equation}
Similar to the homogeneous case, the four point functions on the RHS of equation (\ref{dyson2}) can be derived as follows.
\begin{equation*}
\begin{aligned}
<\rho (y,t) \mbox{ }\psi_{\nu} &(x,\sigma,t) \mbox{ } \psi_{\nu'}^{\dagger}(x',\sigma,t')>
=\bigg(<i \pi \nu\mbox{ } \int_{sgn(x)\infty}^{x} dz \mbox{ } \rho_h(y,t)\rho_h(z,t)\\
&-i\pi\mbox{ }\int_{sgn(x)\infty}^{x} dz \mbox{ }\partial_{v_Ft} \int_{sgn(y)\infty}^z  \rho_h(y,t)\rho_h(z',t)dz' \mbox{ }
+2\nu \lambda \int^{x}_{sgn(x)\infty}dz\mbox{   }\rho_h(y,t)\rho_h(-z,t)>\\
&-<i \pi \nu'\mbox{ } \int_{sgn(x')\infty}^{x'} dz \mbox{ } \rho_h(y,t)\rho_h(z,t')
-i\pi\mbox{ }\int_{sgn(x')\infty}^{x'} dz \mbox{ }\partial_{v_Ft'} \int_{sgn(y)\infty}^z  \rho_h(y,t)\rho_h(z',t')dz' \mbox{ }\\
&+2\nu' \lambda' \int^{x'}_{sgn(x')\infty}dz\mbox{   }\rho_h(y,t)\rho_h(-z,t')>\Bigg)
<\psi_{\nu}(x,\sigma,t)\mbox{ } \psi_{\nu'}^{\dagger}(x',\sigma,t')>
\end{aligned}
\end{equation*}\normalsize
Using the density density correlation functions from equation (\ref{RHOSRHOSNCBT}) and then inserting the above four point function into the RHS of equation (\ref{dyson2}), the necessary Schwinger Dyson equations are obtained. We take into account three cases, viz. RR same side (of the origin), RL same side and RR opposite sides. The remaining three cases: LL same side, LR same side and LL opposite sides are very similar to the former three cases respectively. The   choice of $\lambda,\lambda'$ is discussed in an earlier work \cite{das2018quantum} and also explicitly given in Table \ref{lambdatable}  for the cases discussed.
\begin{table}[h!]
\caption{Choice of  $ \lambda, \lambda'$ for different cases of Green functions.}
{ \begin{tabular}{c  c  c }
\hline\noalign{\smallskip}
 Green's function part   \mbox{  }& $\lambda $ & \hspace{1 cm}$\lambda'$   \\
   \hline
 $\substack{RR \mbox{ }same side  }$ \hspace{1 cm}&0 \hspace{1 cm}&\hspace{1 cm}0   \\[5pt]
$\substack{ RL_1 \mbox{ } same side}$   & $ 1\mbox{  } $ & \hspace{1 cm}0   \\[5pt]
  $\substack{ RL_2 \mbox{ } same side}$   & $ 0\mbox{  } $ & \hspace{1 cm}1   \\[5pt]
  $\substack{ RR_1 \mbox{ } opposite sides}$   & $ 1\mbox{  } $ & \hspace{1 cm}0   \\[5pt]
  $\substack{ RR_2 \mbox{ } opposite sides}$   & $ 0\mbox{  } $ & \hspace{1 cm}1   \\[5pt]
\noalign{\smallskip}\hline
\end{tabular}}
\label{lambdatable}
\end{table}

\subsubsection{RR same side}
\noindent Equation of motion: (here $Z_h = \frac{v_h |R|^2 }{     v_h -  |R|^2 (v_h-v_F)    }$)
\begin{equation}
\begin{aligned}
(i\partial_t + i  v_F   \partial_x) & <T\mbox{  } \psi_{R}(x,\sigma ,t)\psi^{\dagger}_{R}(x',\sigma,t') >
=v_0 \frac{i}{4 \pi   v_h^2}  \mbox{   }( -\frac{2 v_F Z_h  \left(v_F  ( x+x' )   +v_h^2 (t'-t)
\right)}{  (  x  +   x'  +v_h (t-t')) ( x  +  x'  +v_h (t'-t))}\\
&+\frac{v_h (v_h-v_F)}{-v_h(t-t')-x+x'}
-\frac{v_h (v_F+v_h)}{v_h(t-t')-x+x'}
+\frac{v_F^2 Z_h}{x} )
\mbox{   } <T\mbox{  } \psi_{R}(x,\sigma ,t)\psi^{\dagger}_{R}(x',\sigma,t') >
\label{RRss}
\end{aligned}
\end{equation}\normalsize
Substituting the two point function from equation (\ref{SS}),
\small
\begin{equation*}
\begin{aligned}
-i &\bigg(\frac{P (v_h-v_F)}{v_h(t-t')-x+x'}+\frac{Q (v_F+v_h)}{v_h(t-t')+x-x'}
+\frac{X (v_F-v_h)}{-v_h(t-t')+x+x'}+\frac{X (v_F+v_h)}{v_h(t-t')+x+x'}-\frac{\gamma_1 v_F}{x}\bigg) \\
=&v_0 \frac{i}{4 \pi   v_h^2}  \mbox{   }( -\frac{2 v_F Z_h  \left(v_F  ( x+x' )   +v_h^2 (t'-t)
\right)}{  (  x  +   x'  +v_h (t-t')) ( x  +  x'  +v_h (t'-t))}
+\frac{v_h (v_h-v_F)}{-v_h(t-t')-x+x'}
-\frac{v_h (v_F+v_h)}{v_h(t-t')-x+x'}+\frac{v_F^2 Z_h}{x} )
\end{aligned}
\end{equation*}
\normalsize
Upon inserting the explicit expressions of the anomalous exponents from  equations (\ref{luttingerexponents}) and (\ref{le_remaining}) into the above equation, one obtains an identity.


\subsubsection{RL same side}
\noindent Now for this case we have,
\begin{equation*}
\begin{aligned}
 <T\mbox{  } \psi_{R}(x,\sigma ,t)&\psi^{\dagger}_{L}(x',\sigma,t') >
=  <T\mbox{  } \psi_{R}(x,\sigma ,t)\psi^{\dagger}_{L}(x',\sigma,t') >_1
+  <T\mbox{  } \psi_{R}(x,\sigma ,t)\psi^{\dagger}_{L}(x',\sigma,t') >_2
\end{aligned}
\end{equation*}
Equations of motion are:\small
\begin{equation}
\begin{aligned}
(-i\partial_{t'} +& i  v_F   \partial_{x'}) <T\mbox{  }
 \psi_{R}(x,\sigma ,t)\psi^{\dagger}_{L}(x',\sigma,t') >_1\\
  =&v_0(-\frac{i v_F Z_h \left(-v_F  (x' + x)- v_h^2 (t'-t)\right)}
{2 \pi  v_h^2 (-  x  -  x'  +v_h(t-t')) (  x  +  x'  +v_h(t-t'))}
+\frac{i v_F}{2 \pi  v_h (-v_h (t-t')+x+x')}+\frac{i v_F}{2 \pi  v_h (v_h (t-t')+x+x')}\\
-&\frac{i (v_h-v_F)}{4 \pi  v_h (v_h(t-t')+x-x')}-\frac{i (v_h+v_F)}{4 \pi  v_h (v_h(t-t')-x+x')}
+\frac{i v_F^2 Z_h}{4 \pi  v_h^2 x'})\mbox{   } <T\mbox{  } \psi_{R}(x,\sigma ,t)\psi^{\dagger}_{L}(x',\sigma,t') >_1
\label{RLss1}
\end{aligned}
\end{equation}\normalsize
and
\small
\begin{equation}
\begin{aligned}
(i\partial_t + & i  v_F   \partial_x) <T\mbox{  }
 \psi_{R}(x,\sigma ,t)\psi^{\dagger}_{L}(x',\sigma,t') >_2 \\
 =&v_0 (\frac{i v_F Z_h \left(v_F  (x' + x)+ v_h^2 (t'-t)\right)}
{2 \pi  v_h^2 (-  x  -  x'  +v_h(t-t')) (  x  +  x'  +v_h(t-t'))}
+\frac{i v_F}{2 \pi  v_h (v_h (t-t')+x+x')}+\frac{i v_F}{2 \pi  v_h (v_h (t'-t)+x+x')}\\
-&\frac{i (v_F+v_h)}{4 \pi  v_h (v_h(t-t')+x-x')}-\frac{i (v_h-v_F)}{4 \pi  v_h (v_h(t-t')-x+x')}
+\frac{i v_F^2 Z_h}{4 \pi  v_h^2 x})\mbox{   } <T\mbox{  } \psi_{R}(x,\sigma ,t)\psi^{\dagger}_{L}(x',\sigma,t') >_2
\label{RLss2}
\end{aligned}
\end{equation}\normalsize
Substituting the two point functions from equation (\ref{SS}),
\begin{equation*}\small
\begin{aligned}
i (&\frac{ - S (v_F+v_h)}{v_h(t-t')-x+x'}+\frac{  S (v_F-v_h)}
{v_h(t-t')+x-x'}
-\frac{  Y (v_F-v_h)}{-v_h(t-t')+x+x'}-\frac{  Z (v_F+v_h)}{v_h(t-t')+x+x'}
+\frac{ \gamma_1 v_F }{x'}) \\
&\hspace{1.5 cm}=
v_0 (-\frac{i v_F Z_h \left(-v_F  (x' + x)- v_h^2 (t'-t)\right)}
{2 \pi  v_h^2 (-  x  -  x'  +v_h(t-t')) (  x  +  x'  +v_h(t-t'))}+\frac{i v_F^2 Z_h}{4 \pi  v_h^2 x'}
+\frac{i v_F}{2 \pi  v_h (-v_h (t-t')+x+x')}\\
&\hspace{1.5 cm}+\frac{i v_F}{2 \pi  v_h (v_h (t-t')+x+x')}
-\frac{i (v_h-v_F)}{4 \pi  v_h (v_h(t-t')+x-x')}-\frac{i (v_h+v_F)}{4 \pi  v_h (v_h(t-t')-x+x')})\\\\
\end{aligned}
\end{equation*}

\begin{equation*}
\begin{aligned}
i (&\frac{  S (v_F-v_h)}{v_h(t-t')-x+x'}+\frac{  S (v_F+v_h)}{-v_h(t-t')-x+x'}
-\frac{  Y (v_F-v_h)}{-v_h(t-t')+x+x'}-\frac{  Z (v_F+v_h)}{v_h(t-t')+x+x'}
+\frac{ \gamma_1 v_F }{x}) \\
&\hspace{1.34 cm}=
v_0 (\frac{i v_F Z_h \left(v_F  (x' + x)+ v_h^2 (t'-t)\right)}
{2 \pi  v_h^2 (-  x  -  x'  +v_h(t-t')) (  x  +  x'  +v_h(t-t'))} +\frac{i v_F^2 Z_h}{4 \pi  v_h^2 x}
+\frac{i v_F}{2 \pi  v_h (v_h (t-t')+x+x')}\\
&\hspace{1.34 cm}+\frac{i v_F}{2 \pi  v_h (v_h (t'-t)+x+x')}
-\frac{i (v_F+v_h)}{4 \pi  v_h (v_h(t-t')+x-x')}-\frac{i (v_h-v_F)}{4 \pi  v_h (v_h(t-t')-x+x')})
\end{aligned}
\end{equation*}
\normalsize
Upon inserting the explicit expressions of the anomalous exponents from  equations (\ref{luttingerexponents}) and (\ref{le_remaining}), both the above equations are satisfied.

\subsubsection{RR opposite side}
\noindent Now for this case we again have,
\begin{equation*}\small
\begin{aligned}
 <T\mbox{  } \psi_{R}(x,\sigma ,t)&\psi^{\dagger}_{R}(x',\sigma,t') >
=  <T\mbox{  } \psi_{R}(x,\sigma ,t)\psi^{\dagger}_{R}(x',\sigma,t') >_1
+  <T\mbox{  } \psi_{R}(x,\sigma ,t)\psi^{\dagger}_{R}(x',\sigma,t') >_2
\end{aligned}
\end{equation*}
Equations of motion are:
\begin{equation}\small
\begin{aligned}
(-i \partial_{t'} -& i v_F \partial_{x'}) <T\mbox{  } \psi_{R}(x,\sigma ,t)\psi^{\dagger}_{R}(x',\sigma,t') >_1\\
= & v_0  (  \frac{ i v_F Z_h  \left(v_F  (x - x')-v_h^2 (t-t')\right)}{2\pi v_h^2 (-x + x'+v_h (t-t')) (- x + x' +v_h (t'-t))}
 - \frac{i(v_h-v_F)}{4\pi v_h(v_h(t-t')+x-x')} - \frac{i(v_F+v_h)}{4\pi v_h(v_h(t-t')-x+x')} \\
+ &\mbox{  } \frac{iv_F}{2\pi v_h(-v_h(t'-t)+x+x')}+\frac{iv_F}{2\pi v_h(v_h(t'-t)+x+x')}
- \frac{ i v^2_F Z_h}{4 \pi  v_h^2 x'}) <T\mbox{  } \psi_{R}(x,\sigma ,t)\psi^{\dagger}_{R}(x',\sigma,t') >_1 \\
\label{RRos1}
\end{aligned}
\end{equation}\normalsize
and
\small
\begin{equation}
\begin{aligned}
(i\partial_t +& i  v_F   \partial_x) <T\mbox{  } \psi_{R}(x,\sigma ,t)\psi^{\dagger}_{R}(x',\sigma,t') >_2\\
=&v_0(-\frac{i v_F Z_h   \left(v_F (  x -   x' )-v_h^2  (t-t')\right)}{2 \pi  v_h^2 (-  x +   x' +v_h(t-t')) (  x -   x' +v_h(t-t'))}
   -\frac{i v_F}{2 \pi  v_h (-v_h(t-t')+x+x')}-\frac{i v_F}{2 \pi  v_h (v_h(t-t')+x+x')}\\
-&\frac{i (v_h-v_F)}{4 \pi  v_h (v_h(t-t')+x-x')}-\frac{i (v_F+v_h)}{4 \pi  v_h (v_h(t-t')-x+x')}
  + \frac{i v_F^2 Z_h}{4 \pi  v_h^2 x}) <T\mbox{  } \psi_{R}(x,\sigma ,t)\psi^{\dagger}_{R}(x',\sigma,t') >_2
   \label{RRos2}
\end{aligned}
\end{equation}\normalsize
Substituting the two point functions from equation (\ref{OS}),
\begin{equation*}\footnotesize
\begin{aligned}
i( &\frac{  A (v_F-v_h)}{v_h(t-t')-x+x'}+\frac{  B (v_F+v_h)}{-v_h(t-t')-x+x'}
+\frac{ C (v_F+v_h)}{-v_h(t-t')+x+x'}+\frac{  D (v_F-v_h)}{v_h(t-t')+x+x'}-\frac{  \gamma_1 v_F}{x'}) \\
&=  v_0  (  \frac{ i v_F Z_h  \left(v_F  (x - x')-v_h^2 (t-t')\right)}{2\pi v_h^2 (-x + x'+v_h (t-t')) (- x + x' +v_h (t'-t))}- \frac{ i v^2_F Z_h}{4 \pi  v_h^2 x'}
 - \frac{i(v_h-v_F)}{4\pi v_h(v_h(t-t')+x-x')} - \frac{i(v_F+v_h)}{4\pi v_h(v_h(t-t')-x+x')} \\
&+ \mbox{  } \frac{iv_F}{2\pi v_h(-v_h(t'-t)+x+x')}+\frac{iv_F}{2\pi v_h(v_h(t'-t)+x+x')}) \\\\
\end{aligned}
\end{equation*}
\footnotesize
\begin{equation*}
\begin{aligned}
i(& \frac{  A (v_F-v_h)}{v_h(t-t')-x+x'}+\frac{  B (v_F+v_h)}{-v_h(t-t')-x+x'}
-\frac{  C (v_F+v_h)}{v_h(t-t')+x+x'}-\frac{ D (v_F-v_h)}{-v_h(t-t')+x+x'}+\frac{  \gamma_1 v_F}{x}) \\
&=v_0 (-\frac{i v_F Z_h   \left(v_F (  x -   x' )-v_h^2  (t-t')\right)}{2 \pi  v_h^2 (-  x +   x' +v_h(t-t')) (  x -   x' +v_h(t-t'))}+\frac{i v_F^2 Z_h}{4 \pi  v_h^2 x}
   -\frac{i v_F}{2 \pi  v_h (-v_h(t-t')+x+x')}-\frac{i v_F}{2 \pi  v_h (v_h(t-t')+x+x')}\\
&-\frac{i (v_h-v_F)}{4 \pi  v_h (v_h(t-t')+x-x')}-\frac{i (v_F+v_h)}{4 \pi  v_h (v_h(t-t')-x+x')}) \\\\
\end{aligned}
\end{equation*}\normalsize
Upon inserting the explicit expressions of the anomalous exponents from  equations (\ref{luttingerexponents}) and (\ref{le_remaining}), both the above equations are satisfied. Note that the term ($\frac{1}{x+x'}$) in RR opposite sides (equation (\ref{OS})) belongs to the pre-factors which are treated as constants in both sides of the Dyson equation. This is tantamount to the assertion that prefactors are not correctly given by bosonization - only the exponents are.  The prefactors in the present context are spatially dependent and have been adjusted to recover certain limiting behavior - they are to be ignored while one is examining the crucially important Luttinger  exponents. A more convincing way of saying this is - one only looks to equate the time derivative of the logarithms of the two sides of the Schwinger-Dyson equations which forces these prefactors to drop out.

There is one puzzling feature of the arguments that has been presented till now that requires clarification.
Instead of verifying the Schwinger-Dyson equation (SDE) for the Green function (for RR opposite sides and RL same side)  as a whole, we have first written the latter as a sum of two pieces labeled as $ <...>_1 $ and $ <...>_2 $ and verified the SDE that involves  $ (x,t) $ derivatives for the piece labeled $ <....>_2 $ and verified the SDE that involves  $ (x',t') $ derivatives for the piece labeled $ <....>_1 $. This is the same as asserting that the $ (x,t) $ operator (i.e. \small $ i \partial_t + i \nu v_F \partial_x $ \normalsize) acting on $ e^{ i \theta_{\nu}(x,t) } $ behaves as expected but not when it is acting on $ e^{ i \theta_{\nu}(x,t) + 2 \pi i \nu \int^{x} \rho(-y,t) \mbox{  }dy } $. That is, anomalous extensions of the bosonized fields, mandatory for strongly inhomogeneous systems such as the ones being studied here, do not obey the free field equations.  Note that while equating two sides of these equations, the left hand side is purely a power law but for $RR $ opposite sides and $RL $ same side the right hand side has a term of the same functional form as the left hand side plus a term which has a similar but not identical functional form. At the very least we may expect that the two terms whose functional forms match exactly should be equal to each other. It is remarkable that this is indeed the case. Note that for $RR$ and $LL $ same side, the two sides match perfectly without any need for further qualifications. This is important for example to reproduce the dynamical density of states close to the impurity.

It is amply clear and quite remarkable that the explicit formulas for the exponents of the most singular part of the asymptotic Green function of a Luttinger liquid in presence of barriers and wells clustered around an origin as predicted by the non-chiral bosonization method is consistent with the exact Schwinger-Dyson equation of motion for the Green functions. Not only that, the Green functions of NCBT obey the Schwinger-Dyson equation if and only if the anomalous exponents have the precise analytical forms shown in the section  \hyperref[LuttingerExpo]{Anomalous Exponents}. This is a clear vindication of the non-standard harmonic analysis of the density fields and a non-perturbative validation of the NCBT.

\section{ Functional bosonization}

In functional bosonization \cite{fogedby1976correlation,lee1988functional}, one imagines a slowly varying time dependent external potential of the form $ \sum_{q,n} e^{-i q x } e^{ -w_n t} u(q,w_n) $ to be present along with the cluster of barriers and wells. When $ u(q,w_n) \equiv 0 $ and mutual interaction between fermions are absent, the averages are denoted by $ <....>_{SCh} $. It may be shown that the Green function of the system (denoted by $  <T\mbox{  }  \psi_{\nu}(x,t)  \psi^{\dagger}_{\nu'}(x',t')  >_{full}  $) with barriers and wells and including mutual interaction between fermions but without $ u $ , i.e.  $  u(q,w_n) \equiv 0 $  may be obtained by first obtaining the Green function with barriers and wells including $ u $ but without mutual interaction between fermions (denoted by $ <T\mbox{  }  \psi_{\nu}(x,t)  \psi^{\dagger}_{\nu'}(x',t')  > $) and averaging this Green function over $ u $  with a weight $ W[u]  $ as given  below,
\begin{equation}
 <T\mbox{  }  \psi_{\nu}(x,t)  \psi^{\dagger}_{\nu'}(x',t')  >_{full} = \int D[u] \mbox{  }W[u]\mbox{  }  <T\mbox{  }  \psi_{\nu}(x,t)  \psi^{\dagger}_{\nu'}(x',t')  >
\label{FuncB}
\end{equation}
where the weight is,
\begin{equation}
W[u] \equiv \frac{e^{ \frac{ \beta }{2L} \sum_{q,m} \frac{ u(q,w_{m}) u(-q,-w_{m}) }{ v_q} } <TS_U>_{SCh} }{ \int D[u] \mbox{  }e^{ \frac{ \beta }{2L} \sum_{q,m} \frac{ u(q,w_{m}) u(-q,-w_{m}) }{ v_q} } <TS_U>_{SCh} }
\end{equation}
and,
\begin{equation}
<TS_U>_{SCh} = <e^{  -\frac{ \beta }{ L } \sum_{q,m}u(q,w_m)\rho_{q,m} }>_{SCh}
\end{equation}
The nonequilibrium free particle Green function may be related to the equilibrium free particle Green function through the following relation,
\begin{equation}
 <T\mbox{  }  \psi_{\nu}(x,t)  \psi^{\dagger}_{\nu'}(x',t')  >  = \frac{ <TS_U\mbox{  }    \psi_{\nu}(x,t)  \psi^{\dagger}_{\nu'}(x',t')  >_{SCh} }{ <TS_U>_{SCh} }
\end{equation}
This method, like the g-ology methods works well for systems with translational symmetry and can be used to obtain the Green functions of such systems given in equation (\ref{HSS}).  But when applied to systems with impurities, they gives rise to some anomalous terms of the type $\log{[(\frac{x}{v_h}-\frac{x'}{v_F}-(t-t'))]}$ not seen either in conventional perturbation theory nor in the Schwinger Dyson equation. Therefore Green functions obtained are inconsistent with  perturbation theory which makes the study of functional bosonization not suitable for strongly inhomogeneous systems.

\section{ Density matrix renormalization group (DMRG)}
Density matrix renormalization has been the method of choice for the numerical studies in 1D systems. When it comes to calculation of the correlation functions, there are some fundamental differences owing to the energy band structures of the systems. For gapped  systems, the correlation functions decay exponentially with the distance while correlation functions of gapless models decay algebraically with   distance  \cite{eisert2013entanglement}. DMRG is used to optimize the ansatz wavefunctions called as Matrix Product States (MPS) and thus obtaining the correlation functions. But matrix product states are proven useful only for describing the ground states of gapped local Hamiltonians \cite{bultinck2017fermionic,hastings2007entropy,verstraete2008matrix}. Every ground state of a gapped Hamiltonian in 1D can be approximated by a tensor network state to arbitrary precision \cite{hastings2007area}. But M.Andersson et.al. investigated the convergence of DMRG for gapless systems in thermodynamic limit \cite{andersson1999density}. They concluded that when DMRG is used to study a gapless systems of free fermions it gives the wrong particle-hole and density density correlation functions. The expected correlation functions must decay algebraically but the ones obtained from DMRG decay exponentially.

The difficulty in studying gapless systems using DMRG is that  convergence is tough to achieve. Some remedies such as increasing the number of DMRG sweeps, working with finite sized systems are adopted to mitigate the problems. However even if such problems are taken care of, it is less likely that such numerical methods can capture the most singular part of the Green functions which NCBT claims to provide. Nevertheless it will be a challenging problem  for researchers working in the field of DMRG and other numerical techniques to verify the results of NCBT, which are already validated analytically using the Schwinger Dyson equation.


\section{ Conclusion}
In this work, the correlation functions obtained using standard bosonization (g-ology) techniques for homogeneous systems as well as those obtained using the recently developed non-chiral bosonization technique (NCBT) for strongly inhomogeneous systems are validated. Firstly, it has been shown that the correlation functions obey the necessary requirements like current algebra, limiting cases and point  splitting constraints. Secondly, a favorable comparison with the results of standard fermionic perturbation is shown. Thirdly, the Green functions are inserted into the Schwinger-Dyson equation which is the equation of motion of the Green functions resulting in an identity. This serves as a non-perturbative confirmation of these ideas. Lastly, we have discussed that competing analytical approaches such as functional bosonization and numerical ones such as DMRG etc. are not suitable for reproducing the correlation functions   obtained easily by NCBT.

\section*{APPENDICES}

\section*{APPENDIX A: Correlation functions using conventional bosonization}
\label{AppendixA}
\setcounter{equation}{0}
\renewcommand{\theequation}{A.\arabic{equation}}
The conventional bosonization method  that goes by the name `g-ology' can only yield the correlation functions for the extreme cases of homogeneous LL ($|R|=0$) or for a half line  ($|R|=1$).\\

\noindent{\bf Case I: } Green functions for $|R|=0$.\\

\noindent The full Green function is the sum of all the parts. The notion of weak equality is introduced which is denoted by \begin{small} $ A[X_1,X_2] \sim B[X_1,X_2] $ \end{small}. This really means  \begin{small} $ \partial_{t_1} Log[ A[X_1,X_2] ]  = \partial_{t_1} Log[ B[X_1,X_2] ] $\end{small} assuming that A and B do not vanish identically. {\bf Notation:} $X_i \equiv (x_i,\sigma_i,t_i)$ and  $\tau_{12} =  t_1 - t_2$.
\scriptsize
\begin{equation}
\begin{aligned}
\Big\langle T\mbox{  }\psi(X_1)\psi^{\dagger}(X_2) \Big\rangle
=&\Big\langle T\mbox{  }\psi_{R}(X_1)\psi_{R}^{\dagger}(X_2) \Big\rangle +\Big \langle T\mbox{  }\psi_{L}(X_1)\psi_{L}^{\dagger}(X_2)\Big\rangle  \\\\
\end{aligned}
\end{equation}
\normalsize
where
\scriptsize
\begin{equation}
\begin{aligned}
\Big\langle T\mbox{  }\psi&_{R}(X_1)\psi_{R}^{\dagger}(X_2)\Big\rangle \sim
\frac{1}{(x_1-x_2 -v_h \tau_{12})^{P} (-x_1+x_2 -v_h \tau_{12})^{Q} (x_1-x_2 -v_F \tau_{12})^{0.5}} \\
\Big\langle T\mbox{  }\psi&_{L}(X_1)\psi_{L}^{\dagger}(X_2)\Big\rangle \sim
\frac{1}{(x_1-x_2 -v_h \tau_{12})^{Q} (-x_1+x_2 -v_h \tau_{12})^{P}(-x_1+x_2 -v_F \tau_{12})^{0.5}} \\
\label{HSS}
\end{aligned}
\end{equation}\normalsize
and \small
\begin{equation}
P=\frac{(v_h+v_F)^2}{8 v_h v_F} \mbox{ };\mbox{ }Q=\frac{(v_h-v_F)^2}{8 v_h v_F}
\label{expH}
\end{equation}\normalsize
On the other hand, the density-density correlation functions (DDCF) in presence of interactions is given by
\begin{equation}
\begin{aligned}\label{DDCFg}
\langle T \mbox{    }&\rho_s(x_1,\sigma_1,t_1)\rho_s(x_2,\sigma_2,t_2)\rangle
= \frac{1}{4} (\langle T \mbox{    }\rho_h(x_1,t_1)\rho_h(x_2,t_2)\rangle  + \sigma_1 \sigma_2 \langle T \mbox{    }\rho_n(x_1,t_1)\rho_n(x_2,t_2)\rangle )\\
\end{aligned}
\end{equation}
\normalsize
 where $ \rho_h(x,t) =  \rho_s(x,\uparrow,t) + \rho_s(x,\downarrow,t)  $ is the ``holon" density and $ \rho_{n}(x,t) =  \rho_s(x,\uparrow,t) - \rho_{s}(x,\downarrow,t)  $ is the ``spinon" density and
\begin{equation}
\begin{aligned}
\langle T \rho_a(x_1,t_1)\rho_a(x_2,t_2)\rangle  = \frac{v_F  }{ 2\pi^2 v_a } &\mbox{   } \sum_{  \nu = \pm 1 }\bigg (   \frac{-1}{ ( x_1-x_2 + \nu v_a(t_1-t_2) )^2 }
\bigg)
\label{RHOSRHOSg}
\end{aligned}
\end{equation}\normalsize
where $ a = n$ (spinon) or  h (holon).\\\\
\normalsize

\normalsize
\noindent{\bf Case II: } Green functions for $|R|=1$.
\footnotesize
\begin{equation}
\begin{aligned}
\Big\langle T\mbox{  }\psi(X_1)\psi^{\dagger}(X_2) \Big\rangle
=&\Big\langle T\mbox{  }\psi_{R}(X_1)\psi_{R}^{\dagger}(X_2) \Big\rangle +\Big \langle T\mbox{  }\psi_{L}(X_1)\psi_{L}^{\dagger}(X_2)\Big\rangle
+\Big\langle T\mbox{  }\psi_{R}(X_1)\psi_{L}^{\dagger}(X_2) \Big\rangle + \Big\langle T\mbox{  }\psi_{L}(X_1)\psi_{R}^{\dagger}(X_2)\Big\rangle \\\\
\label{breakb}
\end{aligned}
\end{equation}
\normalsize
where
\scriptsize
\begin{equation}
\begin{aligned}
\Big\langle T\mbox{  }\psi&_{R}(X_1)\psi_{R}^{\dagger}(X_2)\Big\rangle \sim
\frac{1}{(x_1-x_2 -v_h \tau_{12})^{P} (-x_1+x_2 -v_h \tau_{12})^{Q} (x_1+x_2 -v_h \tau_{12})^{X} (-x_1-x_2 -v_h \tau_{12})^{X} (x_1-x_2 -v_F \tau_{12})^{0.5}} \\
\Big\langle T\mbox{  }\psi&_{L}(X_1)\psi_{L}^{\dagger}(X_2)\Big\rangle \sim
\frac{1}{(x_1-x_2 -v_h \tau_{12})^{Q} (-x_1+x_2 -v_h \tau_{12})^{P} (x_1+x_2 -v_h \tau_{12})^{X} (-x_1-x_2 -v_h \tau_{12})^{X}(-x_1+x_2 -v_F \tau_{12})^{0.5}} \\
\Big\langle T\mbox{  }\psi&_{R}(X_1)\psi_{L}^{\dagger}(X_2)\Big\rangle \sim
\frac{1}{(x_1-x_2 -v_h \tau_{12})^{X} (-x_1+x_2 -v_h \tau_{12})^{X} (x_1+x_2 -v_h \tau_{12})^{P} (-x_1-x_2 -v_h \tau_{12})^{Q}(x_1+x_2 -v_F \tau_{12})^{0.5}} \\
\Big\langle T\mbox{  }\psi&_{L}(X_1)\psi_{R}^{\dagger}(X_2)\Big\rangle \sim
\frac{1}{(x_1-x_2 -v_h \tau_{12})^{X} (-x_1+x_2 -v_h \tau_{12})^{X}(x_1+x_2 -v_h \tau_{12})^{Q} (-x_1-x_2 -v_h \tau_{12})^{P}(-x_1-x_2 -v_F \tau_{12})^{0.5}} \\
\label{HL}
\end{aligned}
\end{equation}\normalsize
and \small
\begin{equation*}
P=\frac{(v_h+v_F)^2}{8 v_h v_F} \mbox{ };\mbox{ }Q=\frac{(v_h-v_F)^2}{8 v_h v_F} \mbox{ };\mbox{ }  X=\frac{(v_h^2-v_F^2)}{8  v_hv_F}
\end{equation*}

\section*{APPENDIX B:  Correlation functions using NCBT}
\label{AppendixB}
\setcounter{equation}{0}
\renewcommand{\theequation}{B.\arabic{equation}}
\normalsize
The full single particle Green function of a Luttinger liquid in presence of impurities of arbitrary strength ($0<|R|<1$) has been recently derived using the NCBT \cite{das2018quantum}.
The full Green function is the sum of all the parts. The notion of weak equality is introduced which is denoted by \begin{small} $ A[X_1,X_2] \sim B[X_1,X_2] $ \end{small}. This really means  \begin{small} $ \partial_{t_1} Log[ A[X_1,X_2] ]  = \partial_{t_1} Log[ B[X_1,X_2] ] $\end{small} assuming that A and B do not vanish identically. {\bf Notation:} $X_i \equiv (x_i,\sigma_i,t_i)$ and  $\tau_{12} =  t_1 - t_2$.
\footnotesize
\begin{equation}
\begin{aligned}
\Big\langle T\mbox{  }\psi(X_1)\psi^{\dagger}(X_2) \Big\rangle
=&\Big\langle T\mbox{  }\psi_{R}(X_1)\psi_{R}^{\dagger}(X_2) \Big\rangle +\Big \langle T\mbox{  }\psi_{L}(X_1)\psi_{L}^{\dagger}(X_2)\Big\rangle
+\Big\langle T\mbox{  }\psi_{R}(X_1)\psi_{L}^{\dagger}(X_2) \Big\rangle + \Big\langle T\mbox{  }\psi_{L}(X_1)\psi_{R}^{\dagger}(X_2)\Big\rangle \\\\\\\\
\label{break}
\end{aligned}
\end{equation}
\small

\noindent \begin{bf} Case I : $x_1$ and $x_2$ on the same side of the origin\end{bf} \\ \scriptsize

\begin{equation}
\begin{aligned}
\Big\langle T\mbox{  }\psi&_{R}(X_1)\psi_{R}^{\dagger}(X_2)\Big\rangle \sim
\frac{(4x_1x_2)^{\gamma_1}}{(x_1-x_2 -v_h \tau_{12})^{P} (-x_1+x_2 -v_h \tau_{12})^{Q} (x_1+x_2 -v_h \tau_{12})^{X} (-x_1-x_2 -v_h \tau_{12})^{X} (x_1-x_2 -v_F \tau_{12})^{0.5}} \\
\Big\langle T\mbox{  }\psi&_{L}(X_1)\psi_{L}^{\dagger}(X_2)\Big\rangle \sim
\frac{(4x_1x_2)^{\gamma_1}}{(x_1-x_2 -v_h \tau_{12})^{Q} (-x_1+x_2 -v_h \tau_{12})^{P} (x_1+x_2 -v_h \tau_{12})^{X} (-x_1-x_2 -v_h \tau_{12})^{X}(-x_1+x_2 -v_F \tau_{12})^{0.5}} \\
\Big\langle T\mbox{  }\psi&_{R}(X_1)\psi_{L}^{\dagger}(X_2)\Big\rangle \sim
\frac{(2x_1)^{1+\gamma_2}(2x_2)^{\gamma_1}}{2(x_1-x_2 -v_h \tau_{12})^{S} (-x_1+x_2 -v_h \tau_{12})^{S} (x_1+x_2 -v_h \tau_{12})^{Y} (-x_1-x_2 -v_h \tau_{12})^{Z}(x_1+x_2 -v_F \tau_{12})^{0.5}} \\
&\hspace{2cm}+\frac{(2x_1)^{\gamma_1}(2x_2)^{1+\gamma_2}}{2(x_1-x_2 -v_h \tau_{12})^{S} (-x_1+x_2 -v_h \tau_{12})^{S} (x_1+x_2 -v_h \tau_{12})^{Y} (-x_1-x_2 -v_h \tau_{12})^{Z}(x_1+x_2 -v_F \tau_{12})^{0.5}} \\
\Big\langle T\mbox{  }\psi&_{L}(X_1)\psi_{R}^{\dagger}(X_2)\Big\rangle \sim
\frac{(2x_1)^{1+\gamma_2}(2x_2)^{\gamma_1}}{2(x_1-x_2 -v_h \tau_{12})^{S} (-x_1+x_2 -v_h \tau_{12})^{S}(x_1+x_2 -v_h \tau_{12})^{Z} (-x_1-x_2 -v_h \tau_{12})^{Y}(-x_1-x_2 -v_F \tau_{12})^{0.5}} \\
&\hspace{2cm}+\frac{(2x_1)^{\gamma_1}(2x_2)^{1+\gamma_2}}{2(x_1-x_2 -v_h \tau_{12})^{S} (-x_1+x_2 -v_h \tau_{12})^{S}(x_1+x_2 -v_h \tau_{12})^{Z} (-x_1-x_2 -v_h \tau_{12})^{Y}(-x_1-x_2 -v_F \tau_{12})^{0.5}} \\
\label{SS}
\end{aligned}
\end{equation}

\small
\begin{bf}Case II : $x_1$ and $x_2$ on opposite sides of the origin\end{bf} \\ \scriptsize

\begin{equation}
\begin{aligned}
\Big\langle T\mbox{  }\psi&_{R}(X_1)\psi_{R}^{\dagger}(X_2)\Big\rangle \sim
\frac{(2x_1)^{1+\gamma_2}(2x_2)^{\gamma_1}(x_1+x_2)^{-1}(x_1+x_2 + v_F \tau_{12})^{0.5} }{2(x_1-x_2 -v_h \tau_{12})^{A} (-x_1+x_2 -v_h \tau_{12})^{B}(x_1+x_2 -v_h \tau_{12})^{C} (-x_1-x_2 -v_h \tau_{12})^{D} (x_1-x_2 -v_F \tau_{12})^{0.5}} \\
&\hspace{2cm}+\frac{(2x_1)^{\gamma_1} (2x_2)^{1+\gamma_2}(x_1+x_2)^{-1}(x_1+x_2 - v_F \tau_{12})^{0.5}}{2(x_1-x_2 -v_h \tau_{12})^{A} (-x_1+x_2 -v_h \tau_{12})^{B} (x_1+x_2 -v_h \tau_{12})^{D} (-x_1-x_2 -v_h \tau_{12})^{C} (x_1-x_2 -v_F \tau_{12})^{0.5}} \\
\Big\langle T\mbox{  }\psi&_{L}(X_1)\psi_{L}^{\dagger}(X_2)\Big\rangle \sim
\frac{(2x_1)^{1+\gamma_2}(2x_2)^{\gamma_1}(x_1+x_2)^{-1}(x_1+x_2 - v_F \tau_{12})^{0.5} }{2(x_1-x_2 -v_h \tau_{12})^{B} (-x_1+x_2 -v_h \tau_{12})^{A}(x_1+x_2 -v_h \tau_{12})^{D} (-x_1-x_2 -v_h \tau_{12})^{C} (-x_1+x_2 -v_F \tau_{12})^{0.5}} \\
&\hspace{2cm}+\frac{(2x_1)^{\gamma_1} (2x_2)^{1+\gamma_2}(x_1+x_2)^{-1}(x_1+x_2 + v_F \tau_{12})^{0.5}}{2(x_1-x_2 -v_h \tau_{12})^{B} (-x_1+x_2 -v_h \tau_{12})^{A}(x_1+x_2 -v_h \tau_{12})^{C} (-x_1-x_2 -v_h \tau_{12})^{D} (-x_1+x_2 -v_F \tau_{12})^{0.5}} \\
\Big\langle T\mbox{  }\psi&_{R}(X_1)\psi_{L}^{\dagger}(X_2)\Big\rangle \sim \mbox{ }0\\
\Big\langle T\mbox{  }\psi&_{L}(X_1)\psi_{R}^{\dagger}(X_2)\Big\rangle \sim  \mbox{ }0\\
\label{OS}
\end{aligned}
\end{equation}
\normalsize
where ($|R|^2$ is the reflection coefficient)
\footnotesize
\begin{equation}
Q=\frac{(v_h-v_F)^2}{8 v_h v_F} \mbox{ };\mbox{ }  X=\frac{|R|^2(v_h-v_F)(v_h+v_F)}{8  v_h (v_h-|R|^2 (v_h-v_F))}  \mbox{ };\mbox{ }C=\frac{v_h-v_F}{4v_h}
\label{luttingerexponents}
\end{equation}
\normalsize
The other exponents can be expressed in terms of the above exponents.
\footnotesize
\begin{equation}
\begin{aligned}
&P= \frac{1}{2}+Q  \mbox{ };\hspace{0.8 cm}    S=\frac{Q}{C}( \frac{1}{2}-C)   \mbox{ };\hspace{0.85 cm}      Y=\frac{1}{2}+X-C  ;           \\
& Z=X-C\mbox{ };\hspace{0.8 cm}      A=\frac{1}{2}+Q-X \mbox{ };\hspace{0.8 cm}   B=Q-X  \mbox{ };\hspace{1 cm}   \\
&D=-\frac{1}{2}+C   \mbox{ };\hspace{.6 cm}      \gamma_1=X                \mbox{ };\hspace{1.65 cm}    \gamma_2=-1+X+2C;\\
\label{le_remaining}
\end{aligned}
\end{equation}
\normalsize
On the other hand, the density density correlation functions (DDCF) in presence of interactions is given by
\begin{equation}
\begin{aligned}\label{DDCFNCBT}
\langle T \mbox{    }&\rho_s(x_1,\sigma_1,t_1)\rho_s(x_2,\sigma_2,t_2)\rangle
= \frac{1}{4} (\langle T \mbox{    }\rho_h(x_1,t_1)\rho_h(x_2,t_2)\rangle  + \sigma_1 \sigma_2 \langle T \mbox{    }\rho_n(x_1,t_1)\rho_n(x_2,t_2)\rangle )\\
\end{aligned}
\end{equation}
\normalsize
 where $ \rho_h(x,t) =  \rho_s(x,\uparrow,t) + \rho_s(x,\downarrow,t)  $ is the ``holon" density and $ \rho_{n}(x,t) =  \rho_s(x,\uparrow,t) - \rho_{s}(x,\downarrow,t)  $ is the ``spinon" density and
\scriptsize
\begin{equation}
\begin{aligned}
\langle T \rho_a(x_1,t_1)\rho_a(x_2,t_2)\rangle  = \frac{v_F  }{ 2\pi^2 v_a } &\mbox{   } \sum_{  \nu = \pm 1 }\bigg (   \frac{-1}{ ( x_1-x_2 + \nu v_a(t_1-t_2) )^2 }
-\frac{ |R|^2 }{    \bigg( 1 - \delta_{a,h} \frac{(v_h-v_F)}{ v_h } |R|^2   \bigg) }	  \frac{\frac{v_F }{v_a}  \mbox{    } \text{sgn}(x_1) \text{sgn}(x_2)\mbox{   }}{  ( | x_1|+|x_2 | + \nu v_a(t_1-t_2) )^2 }
\bigg)
\label{RHOSRHOSNCBT}
\end{aligned}
\end{equation}\normalsize
where $ a = n$ (spinon) or  h (holon).

\section*{APPENDIX C:  Fermi Bose Correspondence}
\label{AppendixC}
\setcounter{equation}{0}
\renewcommand{\theequation}{C.\arabic{equation}}

The Fermi Bose correspondence is given by equation (\ref{FERBOSE}) of the main text as follows.
\begin{equation}
\psi_k(x) = e^{ i \pi \sum_{ l < k } \int^{\infty}_{ -\infty } dy \mbox{  } \rho_l(y) }  \frac{1}{ \sqrt{ N^{0} } }\sum_{p} n_F(p) \mbox{  } e^{ i \xi(p) }\mbox{   }e^{ i \pi sgn(p)\int^{x}_{ sgn(x)\infty } dy \mbox{  } \rho_k(y) } e^{ -i \pi_k(x) } \mbox{   }\sqrt{ \rho_k(x) }
\label{FERBOSE1}
\end{equation}
where $ n_F(p) = \theta(k_F-|p|) $ and $ N^{0 } = \sum_{p} n_F(p) $.\\\\\normalsize

\noindent{\bf{Theorem:}} The correspondence equation (\ref{FERBOSE1}) is compatible with fermion commutation rules equation (\ref{FERCOM}) in conjunction with equation (\ref{BOSECOM}).\\ \mbox{  }  \\
{\bf{Proof:}} Observe,
\begin{equation*}
\begin{aligned}
&
\psi_j(x) = e^{ i \pi \sum_{ l < j } \int^{\infty}_{ -\infty } dy \mbox{  } \rho_l(y) }  \frac{1}{ \sqrt{ N^{0} } }\sum_{p} n_F(p) \mbox{  } e^{ i \xi(p) }\mbox{   }
e^{ i \pi sgn(p)\int^{x}_{ sgn(x)\infty } dy \mbox{  } \rho_j(y) } e^{ -i \pi_j(x) } \mbox{   }\sqrt{ \rho_j(x) }
\\
&\psi_k(x^{'}) = e^{ i \pi \sum_{ l < k } \int^{\infty}_{ -\infty } dy \mbox{  } \rho_l(y) }  \frac{1}{ \sqrt{ N^{0} } }\sum_{p^{'}} n_F(p^{'}) \mbox{  } e^{ i \xi(p^{'}) }\mbox{   }
e^{ i \pi sgn(p^{'})\int^{x^{'}}_{ sgn(x^{'})\infty } dy \mbox{  } \rho_k(y) } e^{ -i \pi_k(x^{'}) } \mbox{   }\sqrt{ \rho_k(x^{'}) }
\\
&\psi^{\dagger}_k(x^{'}) = \sqrt{ \rho_k(x^{'}) }\mbox{   }e^{ i \pi_k(x^{'}) } \frac{1}{ \sqrt{ N^{0} } }\sum_{p^{'}} n_F(p^{'}) \mbox{  } e^{ -i \xi(p^{'}) } e^{ -i \pi sgn(p^{'})\int^{x^{'}}_{ sgn(x^{'})\infty } dy \mbox{  } \rho_k(y) } e^{ -i \pi \sum_{ l < k } \int^{\infty}_{ -\infty } dy \mbox{  } \rho_l(y) }
\\
\end{aligned}
\end{equation*}
This means,

\begin{equation*}
\begin{aligned}
\psi_j(x)\psi_k(x^{'}) =&\mbox{ } e^{ i \pi \sum_{ l < j } \int^{\infty}_{ -\infty } dy \mbox{  } \rho_l(y) }  \frac{1}{ \sqrt{ N^{0} } }\sum_{p} n_F(p) \mbox{  } e^{ i \xi(p) }\mbox{   }
e^{ i \pi sgn(p)\int^{x}_{ sgn(x)\infty } dy \mbox{  } \rho_j(y) } e^{ -i \pi_j(x) } \mbox{   }\sqrt{ \rho_j(x) }
\\&
  e^{ i \pi \sum_{ l < k } \int^{\infty}_{ -\infty } dy \mbox{  } \rho_l(y) }  \frac{1}{ \sqrt{ N^{0} } }\sum_{p^{'}} n_F(p^{'}) \mbox{  } e^{ i \xi(p^{'}) }\mbox{   }
e^{ i \pi sgn(p^{'})\int^{x^{'}}_{ sgn(x^{'})\infty } dy \mbox{  } \rho_k(y) } e^{ -i \pi_k(x^{'}) } \mbox{   }\sqrt{ \rho_k(x^{'}) }
\end{aligned}
\end{equation*}

\scriptsize
\begin{equation*}
\begin{aligned}
\psi_j(x)\psi_k(x^{'}) =&
  \frac{1}{ \sqrt{ N^{0} } }\sum_{p} n_F(p) \mbox{  } e^{ i \xi(p) }\mbox{   } \frac{1}{ \sqrt{ N^{0} } }\sum_{p^{'}} n_F(p^{'}) \mbox{  } e^{ i \xi(p^{'}) }\mbox{   }
e^{ i \pi sgn(p)\int^{x}_{ sgn(x)\infty } dy \mbox{  } \rho_j(y) } e^{ i \pi sgn(p^{'})\int^{x^{'}}_{ sgn(x^{'})\infty } dy \mbox{  }\rho_k(y)  }e^{ i \pi \sum_{ l < k } \int^{\infty}_{ -\infty } dy \mbox{  }    \rho_l(y) } \\&
 e^{ i \pi \sum_{ l < j } \int^{\infty}_{ -\infty } dy \mbox{  } \rho_l(y) }
\mbox{   }
\sqrt{ (\rho_j(x) + \delta(0)) ( \rho_k(x^{'}) + \delta_{j,k}\delta(x-x^{'}) + \delta(0)) } e^{ -i \pi_j(x) -i \pi_k(x^{'}) }
    \mbox{   }e^{ i \pi \sum_{ l < k }
   \delta_{j,l}  }\mbox{   }e^{ i \pi sgn(p^{'}) \mbox{  }\delta_{j,k}\theta(x^{'}-x) }\mbox{   }
\end{aligned}
\end{equation*}
\normalsize
Also,

\scriptsize
\begin{equation*}
\begin{aligned}
\psi_k(x^{'})\psi_j(x) =
&
  \frac{1}{ \sqrt{ N^{0} } }\sum_{p^{'}} n_F(p^{'}) \mbox{  } e^{ i \xi(p^{'}) }\mbox{   }
   \frac{1}{ \sqrt{ N^{0} } }\sum_{p} n_F(p) \mbox{  } e^{ i \xi(p) }\mbox{   }
e^{ i \pi sgn(p^{'})\int^{x^{'}}_{ sgn(x^{'})\infty } dy \mbox{  } \rho_k(y) } e^{ i \pi sgn(p)\int^{x}_{ sgn(x)\infty } dy \mbox{  }\rho_j(y)  } e^{ i \pi \sum_{ l < j } \int^{\infty}_{ -\infty } dy \mbox{  }
    \rho_l(y) } \\
&e^{ i \pi \sum_{ l < k } \int^{\infty}_{ -\infty } dy \mbox{  } \rho_l(y) }
\mbox{   }
\sqrt{ (\rho_k(x^{'}) + \delta(0)) ( \rho_j(x) + \delta_{j,k}\delta(x-x^{'}) + \delta(0)) } e^{ -i \pi_j(x) -i \pi_k(x^{'}) }
    \mbox{   }e^{ i \pi \sum_{ l < j }
   \delta_{k,l}  }\mbox{   }e^{ i \pi sgn(p) \mbox{  }\delta_{j,k}\theta(x-x^{'}) }\mbox{   }
\end{aligned}
\end{equation*}
\normalsize
Now,
\begin{equation*}
\begin{aligned}
(\rho_j(x) + \delta(0)) ( \rho_k(x^{'}) + \delta_{j,k}\delta(x-x^{'}) + \delta(0)) &=
 (\rho_j(x) + \delta(0)) ( \rho_k(x^{'}) + \delta(0))+ \delta_{j,k}\delta(x-x^{'}) (\rho_j(x) + \delta(0))\\
 & =
(  \rho_k(x^{'})+ \delta(0)) (\rho_j(x) + \delta_{j,k}\delta(x-x^{'}) + \delta(0))
\\& =
 ( \rho_k(x^{'}) + \delta(0)) (\rho_j(x) + \delta(0))+ \delta_{j,k}\delta(x-x^{'})  ( \rho_k(x^{'}) + \delta(0))
\end{aligned}
\end{equation*}
\normalsize
Hence,
\footnotesize
\begin{equation*}
\begin{aligned}
\{\psi_j(x),\psi_k(x^{'}) \} =&
  \frac{1}{ \sqrt{ N^{0} } }\sum_{p^{'}} n_F(p^{'}) \mbox{  } e^{ i \xi(p^{'}) }\mbox{   }
   \frac{1}{ \sqrt{ N^{0} } }\sum_{p} n_F(p) \mbox{  } e^{ i \xi(p) }\mbox{   }
e^{ i \pi sgn(p^{'})\int^{x^{'}}_{ sgn(x^{'})\infty } dy \mbox{  } \rho_k(y) } e^{ i \pi sgn(p)\int^{x}_{ sgn(x)\infty } dy \mbox{  }\rho_j(y)  }
\\&
e^{ i \pi \sum_{ l < j } \int^{\infty}_{ -\infty } dy \mbox{  }
    \rho_l(y) } e^{ i \pi \sum_{ l < k } \int^{\infty}_{ -\infty } dy \mbox{  } \rho_l(y) }
\sqrt{ (\rho_k(x^{'}) + \delta(0)) ( \rho_j(x) + \delta_{j,k}\delta(x-x^{'}) + \delta(0)) } e^{ -i \pi_j(x) -i \pi_k(x^{'}) }
 \\&( e^{ i \pi \sum_{ l < k }
   \delta_{j,l}  }\mbox{   }e^{ i \pi sgn(p^{'}) \mbox{  }\delta_{j,k}\theta(x^{'}-x) } +    \mbox{   }e^{ i \pi \sum_{ l < j }
   \delta_{k,l}  }\mbox{   }e^{ i \pi sgn(p) \mbox{  }\delta_{j,k}\theta(x-x^{'}) } )\\
\end{aligned}
\end{equation*}
\normalsize

\noindent But,

\[
 ( e^{ i \pi \sum_{ l < k }
   \delta_{j,l}  }\mbox{   }e^{ i \pi sgn(p^{'}) \mbox{  }\delta_{j,k}\theta(x^{'}-x) } +    \mbox{   }e^{ i \pi \sum_{ l < j }
   \delta_{k,l}  }\mbox{   }e^{ i \pi sgn(p) \mbox{  }\delta_{j,k}\theta(x-x^{'}) } ) = 0
\]
always. Hence,
\[
\boxed{\{\psi_j(x),\psi_k(x^{'}) \} = 0}\\
\]

\noindent Now,

\begin{equation*}
\begin{aligned}
\psi_j(x)\psi^{\dagger}_k(x^{'}) =& e^{ i \pi \sum_{ l < j } \int^{\infty}_{ -\infty } dy \mbox{  } \rho_l(y) }  \frac{1}{ \sqrt{ N^{0} } }\sum_{p} n_F(p) \mbox{  } e^{ i \xi(p) }\mbox{   }
e^{ i \pi sgn(p)\int^{x}_{ sgn(x)\infty } dy \mbox{  } \rho_j(y) } e^{ -i \pi_j(x) } \mbox{   }\sqrt{ \rho_j(x) }
\\
 & \times \sqrt{ \rho_k(x^{'}) }\mbox{   }e^{ i \pi_k(x^{'}) } \frac{1}{ \sqrt{ N^{0} } }\sum_{p^{'}} n_F(p^{'}) \mbox{  } e^{ -i \xi(p^{'}) } e^{ -i \pi sgn(p^{'})\int^{x^{'}}_{ sgn(x^{'})\infty } dy \mbox{  } \rho_k(y) } e^{ -i \pi \sum_{ l < k } \int^{\infty}_{ -\infty } dy \mbox{  } \rho_l(y) }
\end{aligned}
\end{equation*}

\begin{equation*}
\begin{aligned}
\psi^{\dagger}_k(x^{'}) \psi_j(x)  =& \sqrt{ \rho_k(x^{'}) }\mbox{   }e^{ i \pi_k(x^{'}) } \frac{1}{ \sqrt{ N^{0} } }\sum_{p^{'}} n_F(p^{'}) \mbox{  } e^{ -i \xi(p^{'}) } e^{ -i \pi sgn(p^{'})\int^{x^{'}}_{ sgn(x^{'})\infty } dy \mbox{  } \rho_k(y) } e^{ -i \pi \sum_{ l < k } \int^{\infty}_{ -\infty } dy \mbox{  } \rho_l(y) }
\\&
\times e^{ i \pi \sum_{ l < j } \int^{\infty}_{ -\infty } dy \mbox{  } \rho_l(y) }  \frac{1}{ \sqrt{ N^{0} } }\sum_{p} n_F(p) \mbox{  } e^{ i \xi(p) }\mbox{   }
e^{ i \pi sgn(p)\int^{x}_{ sgn(x)\infty } dy \mbox{  } \rho_j(y) } e^{ -i \pi_j(x) } \mbox{   }\sqrt{ \rho_j(x) }
\end{aligned}
\end{equation*}

Or,

\begin{equation*}
\begin{aligned}
\psi_j(x)\psi^{\dagger}_k(x^{'}) = & e^{ -i \pi_j(x) }  e^{ i \pi \sum_{ l < j } \int^{\infty}_{ -\infty } dy \mbox{  }
 \rho_l(y)   } \frac{1}{ \sqrt{ N^{0} } }\sum_{p} n_F(p) \mbox{  } e^{ i \xi(p) }\mbox{   }
e^{ i \pi sgn(p)\int^{x}_{ sgn(x)\infty } dy \mbox{  } \rho_j(y) }\mbox{   }\sqrt{ \rho_j(x) }
\\&
  \times \sqrt{ \rho_k(x^{'}) }\mbox{   } \frac{1}{ \sqrt{ N^{0} } }\sum_{p^{'}} n_F(p^{'}) \mbox{  } e^{ -i \xi(p^{'}) } e^{ -i \pi sgn(p^{'})\int^{x^{'}}_{ sgn(x^{'})\infty } dy \mbox{  } \rho_k(y) } e^{ -i \pi \sum_{ l < k } \int^{\infty}_{ -\infty } dy \mbox{  }
   \rho_l(y) }e^{ i \pi_k(x^{'}) }\\
\psi^{\dagger}_k(x^{'}) \psi_j(x)  =&  e^{ -i \pi_j(x) }\sqrt{  \rho_k(x^{'}) - \delta_{j,k} \delta(x-x^{'}) } \mbox{   } \frac{1}{ \sqrt{ N^{0} } }\sum_{p^{'}} n_F(p^{'}) \mbox{  } e^{ -i \xi(p^{'}) }e^{ -i \pi sgn(p^{'})\int^{x^{'}}_{ sgn(x^{'})\infty } dy \mbox{  } (\rho_k(y) - \delta_{j,k} \delta(x-y))   }
\\&
 e^{ -i \pi \sum_{ l < k } \int^{\infty}_{ -\infty } dy \mbox{  } (\rho_l(y) - \delta_{j,l} \delta(x-y) ) }
\times  e^{ i \pi \sum_{ l < j } \int^{\infty}_{ -\infty } dy \mbox{  } (\rho_l(y) - \delta_{k,l} \delta(x^{'}-y) )  }  \frac{1}{ \sqrt{ N^{0} } }\sum_{p} n_F(p) \mbox{  } \\
&e^{ i \xi(p) }\mbox{   }
e^{ i \pi sgn(p)\int^{x}_{ sgn(x)\infty } dy \mbox{  }(\rho_j(y) - \delta_{k,j} \delta(x^{'}-y) ) }\mbox{   }\sqrt{
( \rho_j(x) - \delta_{k,j} \delta(x^{'}-x) ) }e^{ i \pi_k(x^{'}) }
\end{aligned}
\end{equation*}

Or,

\begin{equation*}
\begin{aligned}
\psi^{\dagger}_k(x^{'}) \psi_j(x)  =&  e^{ -i \pi_j(x) }\frac{1}{ N^{0} }\sum_{p,p^{'}}  n_F(p) n_F(p^{'}) \mbox{  } e^{ -i \xi(p^{'}) } e^{ i \xi(p) }\mbox{   }\sqrt{  \rho_k(x^{'}) - \delta_{j,k} \delta(x-x^{'}) } \mbox{   } \sqrt{
 \rho_j(x) - \delta_{k,j} \delta(x^{'}-x) }
\\&
 e^{ -i \pi sgn(p^{'})\int^{x^{'}}_{ sgn(x^{'})\infty } dy \mbox{  } \rho_k(y)  }
 e^{ -i \pi \sum_{ l < k } \int^{\infty}_{ -\infty } dy \mbox{  } \rho_l(y)  }
  e^{ i \pi \sum_{ l < j } \int^{\infty}_{ -\infty } dy \mbox{  } \rho_l(y)    }\mbox{   }
e^{ i \pi sgn(p)\int^{x}_{ sgn(x)\infty } dy \mbox{  }\rho_j(y)  }\mbox{   }e^{ i \pi_k(x^{'}) }
\\&
\times  e^{ i \pi sgn(p^{'})\mbox{  }\delta_{j,k} \theta(x^{'}-x)  }
 e^{ i \pi \sum_{ l < k } \mbox{  }\delta_{j,l}   } e^{ -i \pi \sum_{ l < j } \mbox{  } \delta_{k,l}  }
e^{ -i \pi sgn(p)\mbox{  } \delta_{k,j} \theta(x-x^{'})  }
\\
\psi_j(x)\psi^{\dagger}_k(x^{'}) =  &e^{ -i \pi_j(x) } \frac{1}{ N^{0} }\sum_{p,p^{'}} n_F(p) n_F(p^{'}) \mbox{  } e^{ -i \xi(p^{'}) }  e^{ i \xi(p) }\mbox{   }\sqrt{ \rho_j(x) }
   \sqrt{ \rho_k(x^{'}) }
\\&
 e^{ i \pi \sum_{ l < j } \int^{\infty}_{ -\infty } dy \mbox{  }
 \rho_l(y)   }
e^{ i \pi sgn(p)\int^{x}_{ sgn(x)\infty } dy \mbox{  } \rho_j(y) }\mbox{   } e^{ -i \pi sgn(p^{'})\int^{x^{'}}_{ sgn(x^{'})\infty } dy \mbox{  } \rho_k(y) } e^{ -i \pi \sum_{ l < k } \int^{\infty}_{ -\infty } dy \mbox{  }
   \rho_l(y) }e^{ i \pi_k(x^{'}) }
\end{aligned}
\end{equation*}
After setting  $ e^{ -i \xi(p^{'}) } e^{ i \xi(p) } = \delta_{p,p^{'}} $ it is possible to conclude that,
\[
\boxed{\{\psi_j(x),\psi^{\dagger}_k(x^{'})\} =  \delta_{j,k} \delta(x-x^{'})}
\]
\section*{APPENDIX D:  Mathematica commands to verify Schwinger Dyson equations}
\label{AppendixD}
\setcounter{equation}{0}
\renewcommand{\theequation}{B.\arabic{equation}}

For verifying any of the equations in Appendix I, define two variables LHS and RHS and type both sides of the equation as follows:
\begin{equation*}
\begin{aligned}
&LHS:=-i   \left( P\frac{  (v_h-v_F) }{  v_h(t-t')-x+x'  } +..... \right)\\
&RHS:= \frac{ v_0\mbox{  } i }{ 4 \pi v_h^2  } \left(\frac{ v_h(v_h-v_F) }{  v_h(t-t')-x+x'  } +.....\right)
\end{aligned}
\end{equation*}

\noindent{\bf Method I:\\} \\
Define all the variables.
\[
Q:= \frac{ (v_h-v_F)^2 }{  8 v_h v_F } \mbox{ };\mbox{ }P:= \frac{1}{2}+Q
\]
and so on from the section ``Anomalous exponents'' in the main text.
Check
\[
LHS-RHS
\]
to get zero.\\

\noindent{\bf Method II:\\} \\
Use the command
\[
\text{SolveAlways}[LHS==RHS,\{x,x',t,t'\}]
\]
\\
This leads to expressions for the anomalous exponents (denoted by upper case letters of the alphabet) which can then be matched with those in the  section ``Anomalous exponents'' in the main text.\\

\normalsize
\noindent{\bf Acknowledgements}\\
Our deepest gratitude to the lead developer of the ITensor: E. Miles Stoudenmire who has always been very helpful to us whenever we approached him with our querries and doubts. A part of this work was done with financial support from Department of Science and Technology, Govt. of India DST/SERC: SR/S2/CMP/46 2009.\\

\noindent{\bf References\\}
\bibliographystyle{iopart-num}
\bibliography{ref}

\end{document}